\documentclass[11pt, a4paper]{article}

\usepackage{amsmath}
\usepackage{amssymb}
\usepackage{amsthm,url}
\usepackage[hmargin=3cm,vmargin=3cm]{geometry}
\usepackage{graphicx}
\usepackage{latexsym}

\newtheorem{Theorem}{Theorem}
\newtheorem{Proposition}[Theorem]{Proposition}
\newtheorem{Corollary}[Theorem]{Corollary}

\newtheorem{Fact}[Theorem]{Fact}
\newtheorem{Lemma}[Theorem]{Lemma}
\theoremstyle{definition}
\newtheorem{Definition}[Theorem]{Definition}
\newtheorem{Example}[Theorem]{Example}
\newtheorem*{OpenQ}{Open Question}

\newcommand{\seq}[1]{\mathbf{#1}}
\newcommand{\floor}[1]{\left\lfloor #1 \right\rfloor}
\newcommand{\ceil}[1]{\left\lceil #1 \right\rceil}
\newcommand{\Beta}{\text{B}}
\newcommand{\N}{\mathbf{N}}

\let\oldmarginpar\marginpar
\renewcommand\marginpar[1]{\-\oldmarginpar[\raggedleft\footnotesize #1]%
{\raggedright\footnotesize #1}}

\begin{document}

\title{Von Neumann Normalisation of a   Quantum Random Number Generator\thanks{An extended abstract has appeared in A. A. Abbott, C. S. Calude. Von Neumann normalisation and symptoms of randomness: An application to sequences of quantum random bits,  in C. S. Calude, J. Kari, I. Petre, G. Rozenberg (eds.). {\em   Proc.  10th International Conference Unconventional Computation},  Lecture Notes  Comput. Sci. 6714, Springer, Heidelberg, 2011, 40--51.}}
\author{Alastair A. Abbott\thanks{AA was in part supported by the CDMTCS and FP7 Grant PIRSES-2010-269151-RANPHYS.}\qquad  Cristian S. Calude\thanks{CC was in part supported by the CDMTCS, UoA R\&SL grant and FP7 Grant PIRSES-2010-269151-RANPHYS.}\\
Department of Computer Science\\
University of Auckland\\
Private Bag 92019, Auckland, New Zealand\\
\url{www.cs.auckland.ac.nz/~{aabb009,cristian}}
}

\date{\today}
\maketitle

\begin{abstract}
In this paper we study von Neumann un-biasing normalisation for
ideal and real quantum random number generators, operating on finite strings or infinite bit sequences. In the ideal cases one can obtain the desired un-biasing. This relies critically on the independence of the source, a notion we rigorously define for our model. In real cases, affected by imperfections in measurement and hardware, one cannot achieve a true un-biasing, but, if the bias  ``drifts sufficiently slowly'', the result can be  arbitrarily close  to un-biasing. For infinite sequences, normalisation can both increase or \emph{decrease} the (algorithmic) randomness of the generated sequences.

A successful application of von Neumann normalisation---in fact, any un-biasing transformation---does exactly what it promises, {\it un-biasing}, 
 one  (among infinitely many) symptoms of randomness; it will not produce ``true'' randomness.

\end{abstract}
\thispagestyle{empty}

\section{Introduction}

The outcome of some individual quantum-mechanical events cannot in principle be predicted, so they are thought of as ideal sources of random numbers. An incomplete list of quantum phenomena used for random number generation include nuclear decay radiation sources~\cite{Schmidt:1970aa}, the quantum mechanical noise in electronic circuits known as shot noise~\cite{Shen:2010vn}, photons travelling through a semi-transparent mirror~\cite{Kwon:2009aa,Pironia:2010aa,Stefanov:2000aa,Svozil:1990aa,Svozil:2009aa} or photon arrival times~\cite{Wayne:2009kx,Stipcevic:2007fk,Y.-Xie:2005uq}. Our methods are primarily developed to address these latter photon-based quantum random number generators (QRNGs), one of the most direct and popular ways to generate QRNs, but many of our mathematical results will be applicable to other QRNGs.

Due to imperfections in measurement and hardware, QRNGs are biased and operate non-independently in their generation of bits, two symptoms of non-randomness~\cite{Calude:2002fk}.\footnote{As discussed in~\cite{Abbott:aa}, ``true randomness'' does not mathematically exist. Various forms of algorithmic randomness~\cite{Downey:2010aa} are each defined by an infinity of conditions, some ``statistical'' (like bias), some ``non-statistical'' (like lack of computable correlations).}   The first and simplest technique for reducing bias  was invented by 
 von Neumann~\cite{Neumann:2012uq}. It considers pairs of bits, and takes one of three actions: a) pairs of equal bits are discarded; b)  the pair  01 becomes 0; c) the pair 10 becomes 1.  Contrary to wide spread claims, the technique  works for some sources of bits, but not for all.
 The source of constantly biased bits is effectively transformed into one in which the probabilities of 0 and 1 are equal:
 50\% for each. As we shall show, a stronger property is true:  the un-biasing  works not only for bits but for all reasonably long bit-strings.
However, if the bias is not constant
the procedure does not work.
Furthermore, the von Neumann procedure cannot assure ``true randomness'' in its output.
We briefly emphasise that bias is a property of the \emph{source} of bits which only gives the expected frequency of 0's and 1's in the strings produced.

To understand the behaviour of QRNGs we need to study the
un-biasing transformations on both (finite) strings and (infinite) sequences of bits produced by the source. 
In this paper we will focus on von Neumann normalisation\footnote{Many improvements of the scheme have been proposed~\cite{Elias:2043aa,Peres:1992aa}.} because
it is very simple, easy to implement, and (along with the more efficient iterated version due to Peres~\cite{Peres:1992aa} for which the results will also apply) is widely used by current proposals for QRNGs~\cite{Kwon:2009aa,Ma:2004aa,Quantique:aa,Stefanov:2000aa}.
Similar or stronger normalisation procedures have been studied under additional assumptions in, for example,  Blum~\cite{Blum:1986fv} (the source is a finite Markov chain) or Santha and Vazirani~\cite{SanthaVazirani-1986} (the source is semi-random) or Vadhan~\cite{Vadhan:2011kx}; such additional  hypotheses are satisfied by some physical sources, like a zener diode, but not necessarily by quantum sources.
The widespread use of von Neumann normalisation, however, warrants a proper understanding of its operational quality when used on non-ideal sources regardless of the existence of other methods.

  The main results of this paper are the following.  In the ``ideal case'', the  von Neumann normalised output 
 of an independent constantly biased QRNG is the probability space of the uniform distribution (un-biasing). This result is true for both for  finite strings and for the infinite sequences produced  by  QRNGs (the QRNG runs indefinitely in the latter case).

It is important to note that independence in the mathematical sense of multiplicity of probabilities is a model intended to correspond to the physical notion of independence of outcomes~\cite{Kac:1959fk}. In order to study the theoretical behaviour of QRNGs, which are based on the {\it assumption of physical independence of measurements}, we must translate this appropriately into our formal model. We carefully define independence of QRNGs to achieve this aim.

As explained above, QRNGs do not operate in ideal conditions.
We develop a model for a real-world QRNG in which the bias, rather than holding steady, drifts slowly (within some bounds). 
In this framework we  evaluate  the speed of drift required to be maintained by the source distribution to guarantee that the output distribution is as close as one wishes to the uniform distribution.

We have also examined the effect von Neumann normalisation has on various properties of infinite sequences. In particular, 
Borel normality and (algorithmic) randomness are invariant under normalisation, but for $\varepsilon$-random sequences with $0< \varepsilon < 1$, normalisation can both decrease or increase the randomness of the source.

Finally, we present our results in a mathematical framework which avoids hasty claims which later are disproved.

\section{Notation}\label{sec:notation}
We present the main notation used throughout the paper.

By $2^{X}$ we denote the power set of $X$. By $|X|$ we denote the cardinality of the set of $X$.

Let $B=\{0,1\}$ and  denote by $B^{*}$ the set of all bit-strings ($\lambda$ is the empty string). If $x\in B^{*}$ and $i\in B$ then $|x|$ is the length of $x$ and $\#_i(x)$ represents the number of $i$'s in $x$. By
$B^{n}$ we denote the finite set $\{x\in B^{*}\mid n = |x|\}$. The concatenation product of two subsets $X, Y$ of $B^{*}$ is defined by $XY=\{xy \mid x\in X, y\in Y\}$. If $X=\{x\}$ then
we write $xY$ instead of $\{x\}Y$. By $B^{\omega}$ we denote the set of all infinite binary sequences. For $\seq{x}\in B^{\omega}$
and natural $n$ we denote by $\seq{x}(n)$ the prefix of  $\seq{x}$ of length $n$. We write $w \sqsubset v $ or $w \sqsubset \seq{x}$ in case  $w$ is a prefix of the string $v$ or the sequence
$\seq{x}$.

A prefix-free (Turing) machine is a Turing machine whose domain is
a prefix-free set of strings~\cite{Calude:2002fk}. The  prefix complexity of a string, $H_{W}(\sigma)$,
induced by a prefix-free machine $W$ is
$ H_W(\sigma) = \min \{|p| \; : \; W(p) = \sigma\}. $ 
Fix a computable $\varepsilon$ with $0 < \varepsilon \leq 1$.  An $\varepsilon$--universal prefix-free  machine $U$ is a machine such that for every  machine $W$ there is a constant $c$ (depending on $U$ and $W$) such that $\varepsilon \cdot H_{U}(\sigma) \le H_{W}(\sigma)+c$, for all $\sigma\in B^{*}$. If $\varepsilon=1$ then $U$ is simply called a universal prefix-free  machine. 
 A sequence $\seq{x}\in B^{\omega}$ is called $\varepsilon$--random if
there exists a constant $c$ such that $H_{U}(\seq{x}(n))\ge \varepsilon \cdot n-c$, for all $n\ge 1$. Sequences that are
$1$--random are simply called random.

A sequence $\seq{x}$ is called Borel $ m$--normal $(m \geq 1)$ if for every $1  \leq i \leq 2^{m}$ one has:
$\lim_{n \rightarrow \infty}
N_{i}^{m}(\seq{x }(n))/\lfloor\frac{n}{m}\rfloor = 2^{-m};$
here $N_{i}^{m}(y)$ counts the number of non-overlapping occurrences of the $i$th
(in lexicographical order)
binary string of length $m$ in the string $y$.
The sequence $\seq{x}$ is called Borel normal if
it is Borel $m$--normal, for every natural  $m \geq 1$.

A   probability space is a measure space such that the measure of the whole space is equal to one~\cite{Billingsley:1979aa}.
More precisely, a (Kolmogorov) probability space is a triple  consisting of
a sample space $\Omega$, a
 $\sigma$--algebra $\mathcal{F}$ on $\Omega$, and a  probability measure $P$, i.e.\ a countably additive function defined on $\mathcal{F}$ with values in $[0,1]$ such that $P(\Omega) = 1.$

\section{The finite case}

\subsection{Source probability space and independence}
In this section we define the QRNG source probability space and  the independence
property. \medskip

Consider a string of $n$ independent bits produced by a (biased) QRNG. Let $p_0,p_1$ be the probability that a bit is $0$ or $1$, respectively, with $p_0 + p_1 = 1$, $p_0,p_1 \le 1$. 

	The probability space of bit-strings produced by the QRNG is $(B^n,2^{B^n},P_n)$ where $P_n : 2^{B^n} \to [0,1]$ is defined by  
	
	\begin{equation}
	\label{pn}
	P_n(X) = \sum_{x\in X}p_0^{\#_0(x)}p_1^{\#_1(x)},
	\end{equation}
for all $X \subseteq B^n$.

	\if01
\begin{enumerate}
	\item $P_n(\emptyset)=0$, trivially true;
	\item $P_n(B^n) = 1$, guaranteed by the Binomial Theorem;
	\item For $X,Y\subseteq B^n$, $X\cap Y = \emptyset \implies P_n(X \cup Y) = P_n(X) + P_n(Y)$, trivially true.
\end{enumerate}
\fi

	It is easy to verify that the Kolmogorov axioms are satisfied for the space $(B^n,2^{B^n},P_n)$, so we have:
		
\begin{Fact}\label{prop:finiteProbSpace} The space $(B^n,2^{B^n},P_n)$ with
$P_{n}$ defined in {\rm (\ref{pn})} is a probability space.
\end{Fact}

The space $(B^n,2^{B^n},P_n)$ is just the $n$-fold product of the  single bit probability space $(B,2^B,P_1)$.  For this reason this space is often called  an ``independent identically-distributed bit source''. 
The resulting space is ``independent''  because  each bit is  independent of previous ones. But what is ``an independent probability space''? 
\medskip

Physically the independence of a QRNG  is usually expressed as the impossibility of extracting any information from the flow of bits $x_1, \dots ,x_{k-1}$  to improve chances of predicting the value of $x_k$, other than what one would have from knowing the probability space. The fact that photon-based QRNGs obey this physical independence between photons (and thus generated bits) rather well~\cite{Abbott:2010fk,Stefanov:2000aa} is the primary motivation for our modelling of these devices.  These sources (where the condition of independence still holds) are often termed ``independent-bit sources''~\cite{Vadhan:2011kx}. In a real device we cannot, of course, expect each bit to be identically distributed, so we study this more general case more thoroughly in Section~\ref{sec:varyingBias}.

Formally, two events $A,B\subseteq B^n$ are  independent (in a probability space) if the probability of their intersection coincides with the product of their probabilities~\cite{Busch:1996aa} (a complexity-theoretic approach was developed in~\cite{Calude:2010fk}). This motivates  the definition of independence of a general source probability space given 
in Definition~\ref{defindep}. But first we need the  following simple property:
 
\begin{Fact} For every bit-string $x$ and non-negative integers $n,k$ such that $0\le k+|x| \le n$ we
have:
\begin{equation}
\label{proprod}
P_{n} \left(B^{k}xB^{n-k-|x|}\right) = p_{0}^{\#_{0}(x)} p_{1}^{\#_{1}(x)} = P_{|x|}(\{x\}).
\end{equation}
\end{Fact}

\begin{Definition}\label{defindep}
	The probability space $(B^n,2^{B^n},{\rm Prob}_{n})$ is {\em independent}  if for all $1\le k\le n$ and all $x_1\dots x_k \in B^k$ the events $x_1x_2 \dots x_{k-1} B^{n-k+1}$ and $B^{k-1}x_kB^{n-k}$ are independent, i.e. $${\rm Prob}_{n}\left(x_1 x_2 \dots x_{k-1}x_k B^{n-k}\right) = {\rm Prob}_{n}\left(x_1x_2 \dots x_{k-1} B^{n-k+1}\right) \cdot {\rm Prob}_{n}\left(B^{k-1}x_kB^{n-k}\right).$$
\end{Definition}
\begin{Fact}
The probability space $(B^n,2^{B^n},P_n)$ with $P_{n}$ defined in {\rm (\ref{pn})}   is independent. 
\end{Fact}
\begin{proof}
Using (\ref{proprod}) we have:
\begin{align*}
	P_n\left(x_1 x_2 \dots x_{k-1}x_k B^{n-k}\right) &= p_0^{\#_0(x_1\dots x_k)}p_1^{\#_1(x_1\dots x_k)} \\
	&= p_0^{\#_0(x_1\dots x_{k-1})}p_1^{\#_1(x_1\dots x_{k-1})}p_0^{\#_0(x_k)}p_1^{\#_1(x_k)}\\
	&= P_n\left(x_1x_2 \dots x_{k-1} B^{n-k+1}\right) \cdot P_n\left(B^{k-1}x_kB^{n-k}\right).
\end{align*}
\end{proof}

As we will see later, there are other relevant independent probability spaces.

\subsection{Von Neumann normalisation function}
Here we present formally the von Neumann normalisation procedure.

\medskip

We define the mapping $F : B^2 \to B \cup \{\lambda\}$ as 
\begin{equation*}
	F(x_1 x_2) = 
	\begin{cases}
		\lambda & \text{if $x_1 = x_2$,}\\
		x_1 & \text{if $x_1 \neq x_2$,}
	\end{cases}
\end{equation*}
and $f : B \to B^2$ as $$f(x) = x \bar{x},$$
where $\bar{x}=1-x$. Note that  for all $x \in B$ we have $F(f(x))=x$ and, for all $x_1,x_2\in B$ with $x_1\neq x_2$, $f(F(x_1 x_2))=x_1 x_2$.

\medskip

For $m\le \floor{n/2}$ we define the normalisation function $VN_{n,m} : B^n \to \left(\bigcup_{k\le m}B^k\right) \cup\{\lambda \}$ as
$$VN_{n,m}(x_1 \dots x_n) = F(x_1 x_2)F(x_3 x_4)\cdots F\left(x_{(2\floor{\frac{m}{2}}-1)}x_{2\floor{\frac{m}{2}}}\right).$$

\begin{Fact}
	For all $1< m\le\floor{n/2}$ and $y\in B^m$ there exists an $x \in B^n$ such that $y = VN_{n,m}(x)$.
\end{Fact}
\begin{proof}
	Take $x = f(y_1)f(y_2)\cdots f(y_m)0^{n-2m}$.
\end{proof}

In fact we can define the right inverse normalisation $VN_{n,m}^{-1} : 2^{B^m} \to 2^{B^n}$ as 

\begin{align*}VN_{n,m}^{-1}(Y) = \Bigg\lbrace u_1 f(y_1) u_2 f(y_2)\cdots u_m f(y_m) u_{m+1}v \mid y=y_1\dots y_m \in Y, \\\left. u_i \in \{00,11\}^*, v\in B\cup \{\lambda\}, |v| + 2m + \sum_{i=1}^{m+1} |u_i| = n \right\rbrace ,\end{align*}

\noindent for which  $VN_{n,n}\left( VN_{n,m}^{-1}(y) \right) = \{y\}$ holds for every $y\in B^m$.

\subsection{Target probability space and normalisation}

We now construct the target probability space of the normalised bit-strings over $B^m$ for $m\le \floor{n/2}$, i.e.\ the probability space of the output bit-strings produced by the application of the von Neumann function on the output bit-strings generated by the QRNG.

The von Neumann normalisation function $VN_{n,m}$  transforms
	the source probability space   $(B^n, 2^{B^n}, P_{n})$
	into the target probability space $(B^m, 2^{B^m}, P_{n\to m})$.
	The  target   space of normalised bit-strings of length $1<m\le \floor{n/2}$
	associated to the source probability space  $(B^n, 2^{B^n}, P_{n})$  is the space $(B^m, 2^{B^m}, P_{n\to m})$, where $P_{n\to m} : 2^{B^m} \to [0,1]$ is defined  for all $Y \subseteq B^m$ by the formula:
	
	$$P_{n\to m}(Y) = \frac{P_n\left(VN_{n,m}^{-1}(Y) \right)}{P_n\left(VN_{n,m}^{-1}(B^m) \right)}\raisebox{.9mm}{.}$$
	
	
	\begin{Proposition} The  target   space $(B^m, 2^{B^m}, P_{n\to m})$ of normalised bit-strings of length $1<m\le \floor{n/2}$ associated to the source probability space  $(B^n, 2^{B^n}, P_{n})$
	is a probability space.
\end{Proposition}
\begin{proof}
We need to check only additivity: For $X,Y \subseteq B^m$, $X\cap Y = \emptyset \implies P_{n\to m}(X \cup Y)=P_{n\to m}(X) + P_{n\to m}(Y)$. This equality is valid since $VN_{n,m}^{-1}(X \cup Y) = VN_{n,m}^{-1}(X) \cup VN_{n,m}^{-1}(Y)$ and $P_n\left(VN_{n,m}^{-1}(Y) \cup VN_{n,m}^{-1}(X)\right)=P_n\left(VN_{n,m}^{-1}(Y)\right) + P_n\left(VN_{n,m}^{-1}(X)\right)$, as $VN_{n,m}^{-1}(X) \cap VN_{n,m}^{-1}(Y) = \emptyset$ because $X$ and $Y$ are disjoint.
\end{proof}

\subsection{Normalisation of the output of a source with constant bias}

We now show that von Neumann procedure transforms the source probability space with constant bias into
the probability space with the uniform distribution over $B^m$, i.e.\ the target probability space 
$(B^m, 2^{B^m}, P_{n\to m})$ has $P_{n\to m} = U_{m}$, the uniform distribution. Independence and the constant bias of $P_{n}$ play a crucial role.

\begin{Theorem}[von Neumann]\label{vnfilter}
Assume that $1<m \le \lfloor n/2\rfloor$.	In the target probability space 
$(B^m, 2^{B^m}, P_{n\to m})$ associated to the source probability space 
$(B^n, 2^{B^n}, P_{n})$ 
 we have $P_{n\to m}(Y) = U_m(Y) = |Y| \cdot 2^{-m}$, for every  $Y \subseteq B^m$.
\end{Theorem}
\begin{proof}
	 Since $P_{n\to m}$ is additive it suffices to show that for any $y\in B^m$, $P_{n\to m}(\{y\}) = 2^{-m}$. Let $Z = P_n\left(VN_{n,m}^{-1}(B^m) \right)$.
	 
	 We have (the sums are over all $u_i \in \{00,11\}^*$, $v \in B \cup \{\lambda\}$ such that $|v| + \sum_{i=1}^{m+1}|u_i| = n-2m$):
	\begin{align*}
		P_{n\to m}(\{y\}) &= \frac{1}{Z}\sum_{u_i,v}p_0^{\#_0(u_1f(y_1)\dots u_m f(y_m)u_{m+1}v)}p_1^{\#_1(u_1f(y_1)\dots u_m f(y_m)u_{m+1}v)}\\
		&= \frac{p_0^{\#_0(f(y_1)\dots f(y_m))}p_1^{\#_1(f(y_1)\dots f(y_m))}}{Z}\sum_{u_i,v}p_0^{\#_0(u_1\dots u_{m+1}v)}p_1^{\#_1(u_1\dots u_{m+1}v)}\\
		&= \frac{p_0^m p_1^m}{Z}\sum_{u_i,v}p_0^{\#_0(u_1\dots u_{m+1}v)}p_1^{\#_1(u_1\dots u_{m+1}v)},\\
	\end{align*}
\noindent	which is independent of $y$. Since $P_{n\to m}(B^m) = 1$ and for all $x_1,x_2\in B^m$ we have $P_{n\to m}(\{x_1\}) = P_{n\to m}(\{x_2\})$ it follows that $P_{n\to m}(\{y\}) = 2^{-m} = U_m(\{y\})$; by additivity, for every $Y\subseteq 2^m$ we have $P_{n\to m}(Y) = U_m(Y) = |Y|\cdot 2^{-m}$.
\end{proof}

\medskip

It is natural to check whether the independence and constant bias of the source probability space are essential for the validity of the von Neumann normalisation procedure.

\medskip

\begin{Example} The source probability space $(B^{2}, 2^{B^{2}}, {\rm Prob}_{2})$
where ${\rm Prob}_{2}(00) = 0, {\rm Prob}_{2}(01)= {\rm Prob}_{2}(10)= {\rm Prob}_{2}(11)=1/3$ is independent and ${\rm Prob}_{2\to 1} = U_{1}$.
\end{Example}

\begin{Example} The source probability space $(B^{2}, 2^{B^{2}}, {\rm Prob}_{2})$
where ${\rm Prob}_{2}(00) = {\rm Prob}_{2}(11)= 0, {\rm Prob}_{2}(01)= 1/3,  {\rm Prob}_{2}(10)= 2/3$ is independent but ${\rm Prob}_{2\to 1} \not= U_{1}$.
\end{Example}

\noindent{\bf Comment.} One could present the above examples  in the more general framework
of Theorem~\ref{vnfilter}.

\begin{Theorem} \label{vnalter}
Let $m\ge 1$ and $n=2m$. Consider the source probability space
$(B^{n}, 2^{B^{n}}, {\rm Prob}_{n}) = \Pi_{i=1}^{m} (B^{2}, 2^{B^{2}}, P^{i}_{2})$, where $P^{i}_{2}(01)= P^{i}_{2}(10)$, for all
$1\le i \le m$. Then,  in the target probability space $(B^{m}, 2^{B^{m}}, {\rm Prob}_{n\to m})$,
where ${\rm Prob}_{n} = \Pi_{i=1}^{m}P^{i}_{2}$, we have $ {\rm Prob}_{n\to m}=U_{m}$.\end{Theorem}
\begin{proof} It is easy to check that for every $y=y_{1}\dots y_{m} \in B^{m}$ we have ${\rm Prob}_{n\to m}(\{y_{1}\dots y_{m}\}) = \prod_{i=1}^{m}P^{i}_{2}(y_{i}\bar{y_{i}})/{\rm Prob}_{n}(VN_{n,m}^{-1}(B^m))$, so ${\rm Prob}_{n\to m}(\{y_{1}\dots y_{m}\}) $ does not depend on $y$ (because $P^{i}_{2}(a\bar{a})= P^{i}_{2}(\bar{a}a)$, for every $a\in B$). Hence, $ {\rm Prob}_{n\to m}=U_{m}$.

\end{proof}

The source probability space $(B^{n}, 2^{B^{n}}, {\rm Prob}_{n})$ in Theorem~\ref{vnalter} is not constantly biased and may be independent or not, but von Neumann normalisation still produces the uniform distribution under these conditions.

\begin{Example} The source probability space $(B^{4}, 2^{B^{4}}, {\rm Prob}_{4})$
as in  Theorem~\ref{vnalter}
where $P_{2}^{1} (00) = P_{2}^{1} (01) = 1/3, P_{2}^{1} (10) = 1/4, P_{2}^{1} (11) = 1/12$
and $P_{2}^{2} (00) = 1/12, P_{2}^{2} (01) = 1/4, P_{2}^{2} (10) = P_{2}^{2} (11) = 1/3$
is not independent and ${\rm Prob}_{4\to 2} = U_{2}$.
\end{Example}

The outcome of successive context preparations and measurements, such as is the case for the type of QRNG usually envisioned, are postulated to be independent of previous and future outcomes~\cite{Jauch:1968aa}. This means there must be no causal link between one measurement and the next within the system (preparation and measurement devices included) so that the system has no memory of previous or future events.
For QRNGs this translates into the condition that the probability that each successive bit is either 0 or 1 is independent of the previous bit measured. We will only consider such independent probability spaces, as this is a necessary property of a good RNG, so most QRNGs are designed to conform to this requirement.

The above assumption  needs to be made clear as in high bit-rate experimental configurations to generate QRNs with, e.g.,\ photons, its validity may not always be clear. If the wave-functions of successive photons ``overlap'' the assumption no longer holds and (anti)bunching phenomena may play a role. This is an issue that needs to be more seriously considered in QRNG design and will only become more relevant as the bit-rate of QRNGs is pushed higher and higher.  While we leave study of the nature of these temporal correlations (and any non-independence they may cause) to future research~\cite{Abbott:2010fk}, we pose the following open question which may help to quantify any possible effect they may have.

\begin{OpenQ}
	 Fix an integer $k\ge 0$ and small positive real $\kappa$. Consider the probability space $(B^n,2^{B^n},P_n^\dag)$ where $P_n^\dag$ is a modification of the probability $P_n$ satisfying the conditions that for all $i\le n$ and $x_i \in B$ we have $P_n(B^{i-1}x_i B^{n-i}) = P_n^\dag(B^{i-1}x_i B^{n-i})$, $$\left| P_n^\dag(B^{i-1}x_i B^{n-i}) - P_n^\dag(B^{i-1}x_i B^{n-i}\mid B^{i-k-1}x_{i-k}\dots x_{i-1} B^{n-i-1})\right| \le \kappa,$$ and for all $l > k$ 
\begin{align*}&P_n^\dag(B^{i-1}x_i B^{n-i}\mid B^{i-l-1}x_{i-l}\dots x_{i-1} B^{n-i-1})\\ &= P_n^\dag(B^{i-1}x_i B^{n-i}\mid B^{i-k-1}x_{i-k}\dots x_{i-1} B^{n-i-1}).\end{align*} In other words, the probability of each bit depends on no more than the previous $k$ bits, and the difference in probabilities for a bit between that given by $P_n^\dag$ conditioned on the previous $k$ bits and $P_n$ is no more than $\kappa$. \emph{If the output of such a source is normalised with the von Neumann procedure, how close is the resulting probability space of strings of length $m$ to the uniform distribution (see Definition~\ref{def:vardistance} for a definition of the closeness of probability spaces)?}
\end{OpenQ}
\if01
\begin{proof}
	(forgive the messy notation for now)\\
	For $x\in B^n$ we have 
	\begin{align*}
		P_n^\dag(x) &= p^\dag(x_1)p^\dag(x_2|x_1) \cdots p^\dag(x_n | x_{n-k}\dots x_{n-1})\\
			&= (p+c_1)(p+c_2)\cdots (p+c_n)\text{, where }|c_i|\le [?]\\
			&=p^n \prod_{i=1}^n (1+\frac{c_i}{p})
	\end{align*}
	By Lemma~\ref{lemma:maximiseSumAbsValues}, we know the variation from $P$ will be maximised when each $c_i=\pm \kappa$. The issue is working out what the $[?]$ is for the normalised source.. for un-normalised it is $\kappa$, for normalised it will be some function of $\kappa$. However, this form should still hold, and the results for the constant bias variation can be used to give a result here.
\end{proof}
\fi

\subsection{Normalisation of the output of a source with non-constant bias}\label{sec:varyingBias}

Now we consider the probability distribution obtained if von Neumann normalisation is applied to a string generated from an independent source with a non-constant bias---an ``independent-bit source''. We consider only a bias which varies smoothly; this excludes the effects of sudden noise which could make the bias jump significantly from one bit to the next. 
Such a source corresponds to a QRNG in which the bias varies slowly (drifts) from bit to bit over time, but never too far from its average point. We choose this to model photon-based QRNGs since the primary cause of variation in the bias will be of this nature. For example, the detector efficiencies may vary as a result of slow changes in temperature or power supply. While abrupt changes---which this model does not account for---are plausible, their relatively rare occurrence (in comparison with the bit generation rate in the order of MHz) will mean they have little effect on the resultant distribution.

Let $p_0, p_1 < 1$ and $p_0 + p_1 = 1$ be constant. Let $x=x_{1}x_{2}\ldots x_{n}\in B^n$ be the generated string. Then define the probability of an individual bit $x_{i}$ being either zero or one as
\begin{equation}
\label{qi}
	q_i^{x_i} = 
	\begin{cases}
		p_0 - \varepsilon_i & \text{if $x_i=0$},\\
		p_1 + \varepsilon_i & \text{if $x_i=1$}.
	\end{cases}
\end{equation}
The variation in the bias is bounded, so we require that for all $i$, $$|\varepsilon_i| \le \beta, \mbox{ with } \beta < \min(p_0,p_1).$$ Let $\gamma_i = \varepsilon_{i+1} - \varepsilon_i$.  Furthermore, we assume that the ``speed'' of variation be bounded, i.e.\ there exists a positive $\delta$ such that
\begin{equation}
\label{deltabound}
|\gamma_i| \le \delta,
\end{equation}
 for all $i$. Evidently we have $\delta \le \beta$ (presumably in any real situation $\delta \ll \beta$); however, we introduce two separate constants since they correspond to two physically different (but related) concepts. Note that we will discuss in more detail the importance of these two parameters for the approximation of the uniform distribution and their relevance to calibration of the QRNG later once the analysis is completed. Indeed, the rate of change, $\gamma_i$, is more important; the need for $\beta$ stems from the need to realise that, even though the probabilities can fluctuate, they can only fluctuate in one direction for so long (since $q_i \in [0,1]$), hence  $|\sum_i \gamma_i| = |\varepsilon_n - \varepsilon_1| \le 2\beta$. 

\medskip

For a string $y=y_{1}y_{k}\ldots y_{k}\in B^k$ and positive integer $i$ we introduce,  for convenience, the following notation: $$q_i(y) = q_i^{y_1} q_{i+1}^{y_2} \cdots q_{i+k-1}^{y_k}.$$ 
The following fact will allow us to evaluate the effect of normalisation on such a string.
\begin{Fact}
	\label{fact:diffInProbs}
	The difference in probability between $01$ and $10$ depends only on $\gamma_i$, i.e.  $q_i(01) - q_i(10)=\gamma_i$.
\end{Fact}
\begin{proof}
\begin{align*}
	q_i(01) - q_i(10) &= (p_0 - \varepsilon_i)(p_1 + \varepsilon_{i+1}) - (p_1 + \varepsilon_i)(p_0 - \varepsilon_{i+1})\notag\\
	&= (p_0 + p_1)(\varepsilon_{i+1} - \varepsilon_i)\notag\\
	&= \gamma_i. 
\end{align*}
\end{proof}

Let us first formally define the probability space generated by this QRNG.

\begin{Proposition}\label{prop:varyingBiasSimplePS}
	The probability space of bit-strings produced by the QRNG is $(B^n,2^{B^n},R_n)$ where $R_n : 2^{B^n} \to [0,1]$ is defined  for all $X \subseteq B^n$ as follows:
	\begin{equation}
	\label{rn}
	R_n(X) = \sum_{x\in X}q_1(x).
	\end{equation}
\end{Proposition}
\begin{proof}
We verify only that
\if01
this is valid:
\begin{enumerate}
	\item $R_n(\emptyset)=0$, trivially true;
	\item For $X,Y\subseteq B^n$, $X\cap Y = \emptyset \implies R_n(X \cup Y) = R_n(X) + R_n(Y)$, trivially true;
		\item
		\fi
		 $R_n(B^n) = 1$, which is easily shown since $q_i^0 + q_i^1 = 1$, and $R_n(B^n) = (q_1^0 + q_1^1)\cdots (q_n^0 + q_n^1)$.

\end{proof}

\begin{Fact} 
\label{qrule}
For all $i\ge 1$ and $x,y \in \{0,1\}^{*}$ we have:  $q_{i}(xy) = q_{i}(x)q_{i+|x|}(y).$
\end{Fact}

\begin{Fact} For all $k,n\ge 1$, $x \in \{0,1\}^{*}$ with $0\le k+|x| \le n$ we have:
\begin{equation} 
\label{rproprod}
R_{n}\left(B^{n-k}xB^{n-k-|x|}\right) = q_{n-k+1}(x).
\end{equation}
\end{Fact}
\begin{proof} Using Fact~\ref{qrule} we get:
\begin{align}
	R_{n}\left(B^{n-k}xB^{n-k-|x|}\right) &= \sum_{y\in B^{n-k}} \sum_{z\in B^{n-k-|x|}} q_{1}(yxz)
	\notag\\
	&= \sum_{y\in B^{n-k}} \sum_{z\in B^{n-k-|x|}} q_{1}(y) q_{|y|+1}(x)q_{|y|+|x|+1}(z) \notag\\
	&= q_{n-k+1} (x)  \sum_{y\in B^{n-k}} \sum_{z\in B^{n-k-|x|}}q_{1}(y)q_{|y|+|x|+1}(z) \notag \\
	&= q_{n-k+1} (x)  \sum_{y\in B^{n-k}} q_{1}(y) \left(\sum_{z\in B^{n-k-|x|}}q_{|y|+|x|+1}(z)\right) \notag\\
	&= q_{n-k+1} (x).\notag
\end{align}
\end{proof}

\begin{Fact}
The probability space $(B^n,2^{B^n},R_n)$ with $R_{n}$ defined in {\rm (\ref{rn})}  is independent. 
\end{Fact}
\begin{proof}
Using (\ref{rproprod}) we have:
\begin{align*}
	R_n\left(x_1 x_2 \dots x_{k-1}x_k B^{n-k}\right) &= q_{1}(x_1 x_{2}\dots x_{k-1}x_k) \\
	&= q_{1}(x_1 x_{2}\dots x_{k-1})q_{k}(x_k)\\
	&= R_n\left(x_1x_2 \dots x_{k-1} B^{n-k+1}\right) \cdot R_n \left(B^{k-1}x_kB^{n-k} \right).
\end{align*}
\end{proof}

As with the constantly biased source, we  consider the probability space $R_{n\to m}$. We  first investigate the simplest case $n=2m$. In this situation, for any $y\in B^m$ we have $VN_{n,m}^{-1}(\{y\})=\{f(y_1)f(y_2)\cdots f(y_m)\}$ and $VN_{n,m}^{-1}(B^m) = \{f(z_1)f(z_2)\cdots f(z_m) \mid z=z_1\dots z_m \in B^m\}$.
\begin{Fact}\label{prop:varyingBiasSimpleNormedPS}
	The probability space of normalised bit-strings of length $m=n/2$ is $(B^m,2^{B^m},R_{n\to m})$ where $R_{n\to m} : 2^{B^m} \to [0,1]$ is defined  for all $Y \subseteq B^m$ as follows:
	\begin{equation}
	\label{normalprob}
		R_{n\to m}(Y) = \frac{R_n(VN_{n,m}^{-1}(Y))}{R_n(VN_{n,m}^{-1}(B^m))}\\
		= \sum_{y\in Y} \prod_{i=1}^{m}\frac{q_{2i-1}(f(y_i))}{q_{2i-1}(01) + q_{2i-1}(10)}\raisebox{0.9mm}{.}
	\end{equation}
\end{Fact}
\if01
\begin{enumerate}
	\item $R_{n\to m}(\emptyset)=0$, trivially true;
	\item $R_{n\to m}(B^m) = 1$, trivially true because of the normalisation factors for each bit;
	\item For $X,Y\subseteq B^m$, $X\cap Y = \emptyset \implies R_{n\to m}(X \cup Y) = R_{n\to m}(X) + R_{n\to m}(Y)$, trivially true.
\end{enumerate}
\fi

\subsection{Approximating  the uniform distribution}

Unlike the case for a constantly biased source, we no longer have $q_i(01)=q_i(10)$; from Fact~\ref{fact:diffInProbs} we have $q_i(01)=q_i(10) + \gamma_i$. As a result the normalised equation is no longer the uniform distribution, but only an approximation thereof. We now explore how closely $R_{n\to m}$ approximates $U_m$. 

We first need to define what we mean by approximating $U_m$. 
\begin{Definition}\label{def:vardistance}
	The {\em total variation distance} between two probability measures $P$ and $Q$ over the space $\Omega$ is $\Delta(P,Q)=\max_{A\subseteq \Omega}|P(A) - Q(A)|$. We say that $P$ and $Q$ are $\rho$-close if $\Delta(P,Q) \le \rho$.
\end{Definition}

It is well known (see for example \cite{Vadhan:2011kx}) that
\begin{Lemma}\label{finitedist}
	For finite $\Omega$ we have $\Delta(P,Q)=\frac{1}{2}\sum_{x\in \Omega}|P(\{x\})-Q(\{x\})|$.
\end{Lemma}
The variation $\Delta(R_{n\to m},U_m)$ depends on each $\gamma_i$ and $q_i$ (thus on $p_0$, $p_1$ and each $\varepsilon_i$), but we wish to calculate the worst case in terms of the bounds $\delta, \beta$ and $p_0, p_1$, i.e.\ using Lemma~\ref{finitedist},
\begin{align*}
	\max_{\gamma_i,q_i} \Delta(R_{n\to m},U_m) = \frac{1}{2}\max_{\gamma_i,q_i}\sum_{y\in B^m}|R_{n\to m}(\{y\}) - 2^{-m}|.
\end{align*}
Let us first note that we can write
\begin{align*}
	\frac{q_{2i-1}(f(y_i))}{q_{2i-1}(01) + q_{2i-1}(10)}&=\frac{q_{2i-1}(f(y_i))}{2q_{2i-1}(f(y_i))-(-1)^{y_i}\gamma_{2i-1}}\\
		&=\frac{1}{2}\left(1 + \frac{(-1)^{y_i}\gamma_{2i-1}}{2q_{2i-1}(f(y_i))-(-1)^{y_i}\gamma_{2i-1}} \right)\raisebox{.9mm}{,}
\end{align*}
and hence we have 
$$R_{n\to m}(\{y\})=2^{-m}\prod_{i=1}^m \left(1 + \frac{(-1)^{y_i}\gamma_{2i-1}}{q_{2i-1}(01)+q_{2i-1}(10)} \right)\raisebox{.9mm}{.}$$
We have rewritten the denominator in its original form to emphasise that only the signs $(-1)^{y_i}$ depend on $y$. Thus, we want to find the values of $q_{2i-1}$ and $\gamma_{2i-1}$ which maximise 
\begin{equation}\label{eqn:fnToMax}
	\sum_{y\in B^m}\left|1- \prod_{i=1}^m \left(1 + \frac{(-1)^{y_i}\gamma_{2i-1}}{q_{2i-1}(01)+q_{2i-1}(10)} \right)\right|\raisebox{.9mm}{,}
\end{equation}
subject to the constraints that $|\gamma_\ell| \le \delta$ and $|\varepsilon_\ell| \le \beta$ for $1\le \ell \le n$.

\if03
\begin{Lemma} 
\label{qibound}
For every $j \ge 1$ and $a,b \in \{0,1\}, a\not= b$ we have: $q_{j}^{a}q_{j+1}^{b}
\ge \beta^{2}- \beta + p_{0}p_{1}$, provided $0< \beta < (1 + (1-4p_{0}p_{1})^{1/2})/2$.
\end{Lemma}
\begin{proof} Fix  $j \ge 1$ and note that $\min\{-\varepsilon_{j}, \varepsilon_{j}\} \ge - |\varepsilon_{j} | \ge -\beta$ as $ |\varepsilon_{j} | \le \beta$. Hence  we have:
\begin{align*}
q_{j}^{0}q_{j+1}^{1}	&=  (p_{0} -\varepsilon_{j})(p_{1} + \varepsilon_{j+1})\\
		&\ge p_{0}p_{1}  - p_{0}\beta - p_{1}\beta + \beta^{2}\\
		&\ge p_{0}p_{1}  - \beta  + \beta^{2} >0,
\end{align*}
provided $0< \beta < (1 + (1-4p_{0}p_{1})^{1/2})/2$. Similarly so for $q_{j}^{0}q_{j+1}^{1}$ with $\varepsilon_j$ and $\varepsilon_{j+1}$ swapped.
\end{proof} 
\hline
My concern here is that this bound cannot be obtained. It requires that $\varepsilon_j = -\varepsilon_{j+1} = \pm \beta$. However, we have the constraint that $\varepsilon_{j+1} = \varepsilon_j + \gamma_j$, making this impossible. For $\delta \ll \beta$ (the physical case we want the mathematics to be relevant in) this bound is very optimistic and the real bound is likely very different.
\hline

\begin{Theorem} Assume that $m=n/2$. Consider the probability spaces $(B^n,2^{B^n},R_{n})$ and   $(B^n,2^{B^n},R_{n\to m})$. Assume that $R_{n}$ is given by the functions $q$ given by (\ref{qi})
satisfying the condition (\ref{deltabound}). For every positive real $\rho$, if $\delta < ((1+2\rho)^{1/m}-1)Q$, where  then $\Delta(R_{n},U_{n}) < \rho$.
\end{Theorem}
\begin{proof} Put
\[Q_{m} = \min_{1\le i\le m}(q_{2i-1}^{0}q_{2i}^{1}+q_{2i-1}^{1}q_{2i}^{0})\ge 2(\beta^{2}- \beta + p_{0}p_{1}),\]
\hline
Again my concern is that this bound cannot be nearly obtained. We know $q(01) = q(10)$ iff $\varepsilon_j = \varepsilon_{j+1}$, while $\min(q(01))$ was obtained for $\varepsilon_j = -\varepsilon_{j+1}$.
\hline
 and note that $Q_{m} \ge  2(p_{0}p_{1}  - \beta  + \beta^{2}) >0$ by Lemma~\ref{qibound}. Recall that $\gamma_{i}= \varepsilon_{i+1}-
\varepsilon_{i}$ and,  by (\ref{deltabound}), $|\gamma_{i}|\le \delta$. It is seen that
$\max\{1+ \gamma_{2i-1}/Q_{m}, 1- \gamma_{2i-1}/Q_{m}\} \le 1+ \delta/Q_{m}$. Hence for every $y\in B^{m}$,
\begin{equation}
	R_{n\to m}(\{y\})=2^{-m}\prod_{i=1}^m \left(1 + \frac{(-1)^{y_i}\gamma_{2i-1}}{q_{2i-1}(01)+q_{2i-1}(10)} \right)\le 2^{-m} \left(1+\frac{\delta}{Q_{m}}\right)^{m},
\end{equation}

\noindent hence 
\begin{equation}
	\Delta (R_{n\to m}, U_{m}) \le \frac{1}{2} \left( \left( 1+\frac{\delta}{2( p_{0}p_{1}  - \beta  + \beta^{2})}\right)^{m}-1\right),
\end{equation}
\noindent
so $R_{n\to m}$ and $ U_{m}$ are $\rho$-close provided $\delta < 2((1+2\rho)^{1/m}-1)( p_{0}p_{1}  - \beta  + \beta^{2})$.
\end{proof}
\hline
I have two concerns here. We have $R(y) \le \max_{y,\gamma,q}R(y)$, and $$\max R(y) = 2^{-m}\left( 1+\max_{\gamma,q}\frac{\gamma}{q(01) + q(10)} \right)^m,$$ but it is not the case that $$\max\frac{\gamma}{q(01) + q(10)} = \frac{\delta}{\min( q(01) + q(10))}$$ because $q(01)$ and $q(10)$ depend on $\gamma$ too. Hence again the bound cannot be reached, and (9) should have a strictly less than sign... I am not sure how much larger the derived bound is than the real upper bound on $R(y)$. 

My second concern I feel is more serious. We want to find the $\rho$ such that $\Delta(R,U) \le \rho = \frac{1}{2}\max \sum_y |R(y)-U(y)|$. However, for large enough $m$ we have 
$$\max \sum_y |R(y)-2^{-m}| \ll 2^m |\max R(y) - 2^{-m}|,$$ in fact the bound in equation (10) is unbounded with $m$, whereas it is clear by definition that $\Delta(P,Q) \le 1$ for any probabilities $P,Q$. I thus feel it is important to take into account the summation if we wish to find a reasonable bound on $\rho$. This is why I have taken the approach I did by specifying  Lemma 17 and the derivation after Lemma 18.
\hline
\fi

\begin{Lemma}\label{lemma:maximiseSumAbsValues}
	The function $$g(c_1,\dots,c_n)=\sum_{y\in B^n}\left|\prod_{i=1}^n\left(1+(-1)^{y_i}c_i \right) -1 \right|$$ is strictly increasing for $0\le c_i < 1$, $i=1,\dots,n$ (note that for $1\le i\le n$, $g(c_1,\dots,c_i,\dots,c_n) = g(c_1,\dots,-c_i,\dots,c_n)$).
\end{Lemma}
\begin{proof}
	We take $0\le c_i < 1$ for $1\le i \le n$. For $y=y_1\dots y_n \in B^n$ define $p(y,j)=\prod_{i=1,i\neq j}^n (1+(-1)^{y_i}c_i)$. Without loss of generality pick a $j\le n$ and let $\varepsilon > 0$ be an (arbitrarily small) positive real with $c_j + \varepsilon \le 1$. Note that
	
	$$g(c_1,\dots,c_n)=\sum_{y\in B^{n}}\left|(1+(-1)^{y_j}c_j)p(y,j) -1 \right|.$$ 
	We partition $B^{n}$ as follows:
	\begin{align*}
		Y_1 &= \{ y \mid (1-c_j-\varepsilon)p(y,j) - 1 \ge 0 \},\\
		Y_2 &= \{ y \mid (1-c_j-\varepsilon)p(y,j) - 1 < 0 \text{ and } (1-c_j)p(y,j) - 1 \ge 0 \},\\
		Y_3 &= \{ y \mid (1-c_j)p(y,j) - 1 < 0 \text{ and } (1+c_j)p(y,j) - 1 \ge 0 \},\\
		Y_4 &= \{ y \mid (1+c_j)p(y,j) - 1 < 0 \text{ and } (1+c_j+\varepsilon)p(y,j) - 1 \ge 0 \},\\
		Y_5 &= \{ y \mid (1+c_j+\varepsilon)p(y,j) - 1 < 0 \}.\\
	\end{align*}
	Note that for $y \in B^{n}$, $p(y,j) \ge 0$, and for $y_i \in Y_i$, $i=1,\dots,5$, we have $$p(y_5,j) < p(y_4,j) < p(y_3,j) < p(y_2,j) < p(y_1,j),$$ and $\bigcup_{i=1}^5 Y_i = B^{n}$. We have:
	\begin{align*}
		g(c_1,\dots,c_j+\varepsilon,\dots, c_n) =&\sum_{i=1}^5 \sum_{y \in Y_i}|(1+(-1)^{y_j}c_j + (-1)^{y_j}\varepsilon)p(y,j)-1|\\
		=&\sum_{y\in Y_1}\left[ (1+(-1)^{y_j}c_j)p(y,j)-1 +(-1)^{y_j}\varepsilon p(y,j) \right]\\
		&+\sum_{i=2}^4\sum_{y\in Y_i}(-1)^{y_j}\left[ (1+(-1)^{y_j}c_j)p(y,j)-1 +(-1)^{y_j}\varepsilon p(y,j) \right]\\
		&+\sum_{y\in Y_5}-\left[ (1+(-1)^{y_j}c_j)p(y,j)-1 +(-1)^{y_j}\varepsilon p(y,j) \right]\\
		=&\sum_{i=1}^5 \sum_{y \in Y_i}|(1+(-1)^{y_j}c_j)p(y,j)-1|+2\varepsilon \sum_{i=2}^4\sum_{y\in Y_i}p(y,j)\\
		&-2\sum_{y\in Y_2}\left[(1-c_j)p(y,j)-1 \right]+2\sum_{y\in Y_4}\left[(1+c_j)p(y,j)-1 \right]\\
		=& g(c_1,\dots,c_j,\dots,c_n)+ 2\varepsilon \sum_{y\in Y_3}p(y,j) \\ &- 2\sum_{y\in Y_2}\left[(1-c_j-\varepsilon)p(y,j) - 1 \right]  + 2\sum_{y\in Y_4}\left[(1+c_j+\varepsilon)p(y,j) - 1 \right]\\
		>& g(c_1,\dots,c_j,\dots,c_n),
	\end{align*}
	where the final line follows from the definition of $Y_2$ and $Y_4$. Since this holds for all $j \le n$, $g$ is strictly increasing over $[0,1)^n$.
\end{proof}
Hence in order to maximise \eqref{eqn:fnToMax} we need to maximise the functions
\begin{align}\label{eqn:smallTermToMax}
	 u_j(\varepsilon_j,\gamma_j) = \left|  \frac{\gamma_j}{q_j(01)+q_j(10)} \right| = \left|  \frac{\gamma_j}{(p_0 - \varepsilon_j)(p_1 + \varepsilon_j + \gamma_j) + (p_1 + \varepsilon_j)(p_0 - \varepsilon_j - \gamma_j)} \right|\raisebox{.9mm}{,}
\end{align}
for $j=2i-1$, $1\le i \le m$, subject to the constraints $|\gamma_j| \le \delta$, $|\varepsilon_j| \le \beta$ and $|\varepsilon_{j+1}|=|\varepsilon_j + \gamma_j| \le \beta$.

\begin{Lemma}\label{lemma:maxAlpha}
	For every $j\ge 1$ we have 
	\begin{align}\label{eqn:uMaxConds}
		u_j(\varepsilon_j,\gamma_j) &\le 
		\begin{cases}
			u_j(\beta,-\delta) = u_j(\beta - \delta, \delta) & \text{if $p_1 \ge p_0$,}\\
			u_j(-\beta,\delta) = u_j(-\beta + \delta, -\delta)  & \text{if $p_0 > p_1$,}
		\end{cases}\\
		&= \frac{\delta}{2\left[ p_0 p_1 - \beta(\beta - \delta) - |p_0 - p_1|(\beta - \delta/2) \right]}\raisebox{.9mm}{.}\label{eqn:maxmimisedU}
	\end{align}
\end{Lemma}
\begin{proof}
	We omit the index $j$ as it is not needed in this context. Let $$v(\varepsilon,\gamma)=\frac{\gamma}{(p_0 - \varepsilon)(p_1 + \varepsilon + \gamma) + (p_1 + \varepsilon)(p_0 - \varepsilon - \gamma)}\raisebox{.9mm}{.}$$ Since $q(01)+q(10) > 0$, in order to maximise $u$ we look for maxima and minima of $v$; clearly maxima have $\gamma >0$ and minima have $\gamma < 0$. We use Lagrange multipliers with inequality constraints to find the critical points. We have the following six constraints: $h_1(\varepsilon,\gamma) = \varepsilon - \beta \le 0$, $h_2(\varepsilon,\gamma) = -\varepsilon - \beta \le 0$, $h_3(\varepsilon,\gamma) = \varepsilon + \gamma - \beta \le 0$, $h_4(\varepsilon,\gamma) = -\varepsilon -\gamma - \beta \le 0$, $h_5(\varepsilon,\gamma) = \gamma - \delta \le 0$, $h_6(\varepsilon,\gamma) = -\gamma - \delta \le 0$. We must solve the following equations:
	\begin{align}
		\nabla_{\varepsilon,\gamma}v(\varepsilon,\gamma)+\sum_{i=1}^6\lambda_i \nabla_{\varepsilon,\gamma}h_i(\varepsilon,\gamma)=0,\\
		\lambda_i h_i(\varepsilon,\gamma) = 0\text{\qquad for $i=1,\dots,6$,}\label{eqn:complimentarily}\\
		h_i(\varepsilon,\gamma) \le 0\text{\qquad for $i=1,\dots,6$,}\\
		\begin{cases}
			\lambda_i \ge 0 & \text{for minima, $i=1,\dots,6$,}\\
			\lambda_i \le 0 & \text{for maxima, $i=1,\dots,6$.}
		\end{cases}
	\end{align}
	We say a constraint is inactive if $\lambda_i = 0$ and active otherwise; the condition of complimentarity~\eqref{eqn:complimentarily} captures the notion that a critical point satisfying the constraints either occurs at $h_i(\varepsilon,\gamma) = 0$ or is also a critical point in the unconstrained problem.
	
	Noting that $0< p_0 - \beta \le p_0 + \beta < 1$ and solving, we find the candidate points are: 
	\begin{align*}
		(\varepsilon,\gamma) = 
		\begin{cases}
			(\frac{1}{2}(p_0 - p_1)\pm\frac{\delta}{2}, \mp \delta) &\\
			(\beta, -\delta), (\beta-\delta, \delta) & \text{for $p_0-p_1 \le 2\beta - \delta$,}\\
			(-\beta, \delta), (-\beta + \delta,-\delta) & \text{for $p_1-p_0 \le 2\beta - \delta$.}\\
		\end{cases}
	\end{align*}
	Note that $u(\varepsilon,\gamma)=u(\varepsilon + \gamma, -\gamma)$. Testing values shows the second case maximises $u(\varepsilon,\gamma)$ when $p_1 > p_0$ and the third cases maximises $u(\varepsilon,\gamma)$ for $p_0 > p_1$. For $p_0 = p_1$ both cases give the same value. Substituting in $\varepsilon,\gamma$ and consolidating the cases we arrive at \eqref{eqn:maxmimisedU}.
\end{proof}

\medskip

Next we let $$\alpha=\max_{\gamma_i,\varepsilon_i} u_j(\varepsilon_j,\gamma_j),$$
where $ u_j(\varepsilon_j,\gamma_j)$ comes from (\ref{eqn:smallTermToMax}).

\medskip

Then we have
\begin{align*}
	\max_{\gamma_i,\varepsilon_i}\Delta(R_{n\to m},U_m)&=\frac{1}{2}\sum_{y\in B^m}\left| \prod_{i=1}^m\left( \frac{1}{2}+(-1)^{y_i}\frac{\alpha}{2} \right) - 2^{-m}  \right|\\
	&= \frac{1}{2}\sum_{k=0}^m \binom{m}{k}\left| \left(\frac{1}{2}+\frac{\alpha}{2}\right)^k\left(\frac{1}{2}-\frac{\alpha}{2}\right)^{m-k}-2^{-m}\right|.
\end{align*}
Note that in this worst case, the normalised source acts as an  independent and identically-distributed source with $p_0 = 1/2 \pm \alpha/2$ and the total variation is bounded by that of two binomial sources: one with $p_0 = 1/2$, the other with $p_0=1/2 \pm \alpha/2$ (the number $k$ of successful outcomes is identified with the number of ones in $y$). 

There are two interesting questions: a) what is the quality of the distribution  produced by
a QRNG, i.e.\ how close are $R_{n\to m}$ and $U_m$ in terms of $\alpha$? and b) given a real $\rho \in (0,1)$, how accurate does the QRNG need to be in terms of $\alpha$ to guarantee that $R_{n\to m}$ and $U_m$ are $\rho$ close?

We can take a rough approach to solve the above problems as follows. First note that 

\begin{align*}
	\Delta(R_{n\to m},U_m) &\le \frac{1}{2}\sum_{y\in B^m}\left| \prod_{i=1}^m\left( \frac{1}{2}+(-1)^{y_i}\frac{\alpha}{2} \right) - 2^{-m}  \right|\\
	&\le \frac{1}{2}\sum_{y\in B^m}\frac{1}{2^m} \left( (1+\alpha)^m - 1 \right)\\
	&= \frac{1}{2}\left( (1+\alpha)^m - 1 \right).\\
\end{align*}
So given $\alpha$, $R_{n\to m}$ and $U_m$ are at most $\frac{1}{2}\left( (1+\alpha)^m - 1 \right)$-close. Conversely, $R_{n\to m}$ and $U_m$ are
$\rho$ close if 
\begin{equation}\label{eqn:rhoCloseNaive}
	\alpha \le (1+2\rho )^{1/m} - 1.
\end{equation}
We will express further results in the latter form, focusing on question b), although both are important questions depending on the operational circumstances and results can easily be transformed from one form to the other.

So, by making $\alpha$ very small, $R_{n\to m}$ can be made as close as we wish to the uniform distribution. This is intuitive since $\alpha \to 0$ only as $\delta \to 0$ and we approach the constantly biased source situation. 

\begin{figure}[h]
	\centering
	\includegraphics[scale=0.65]{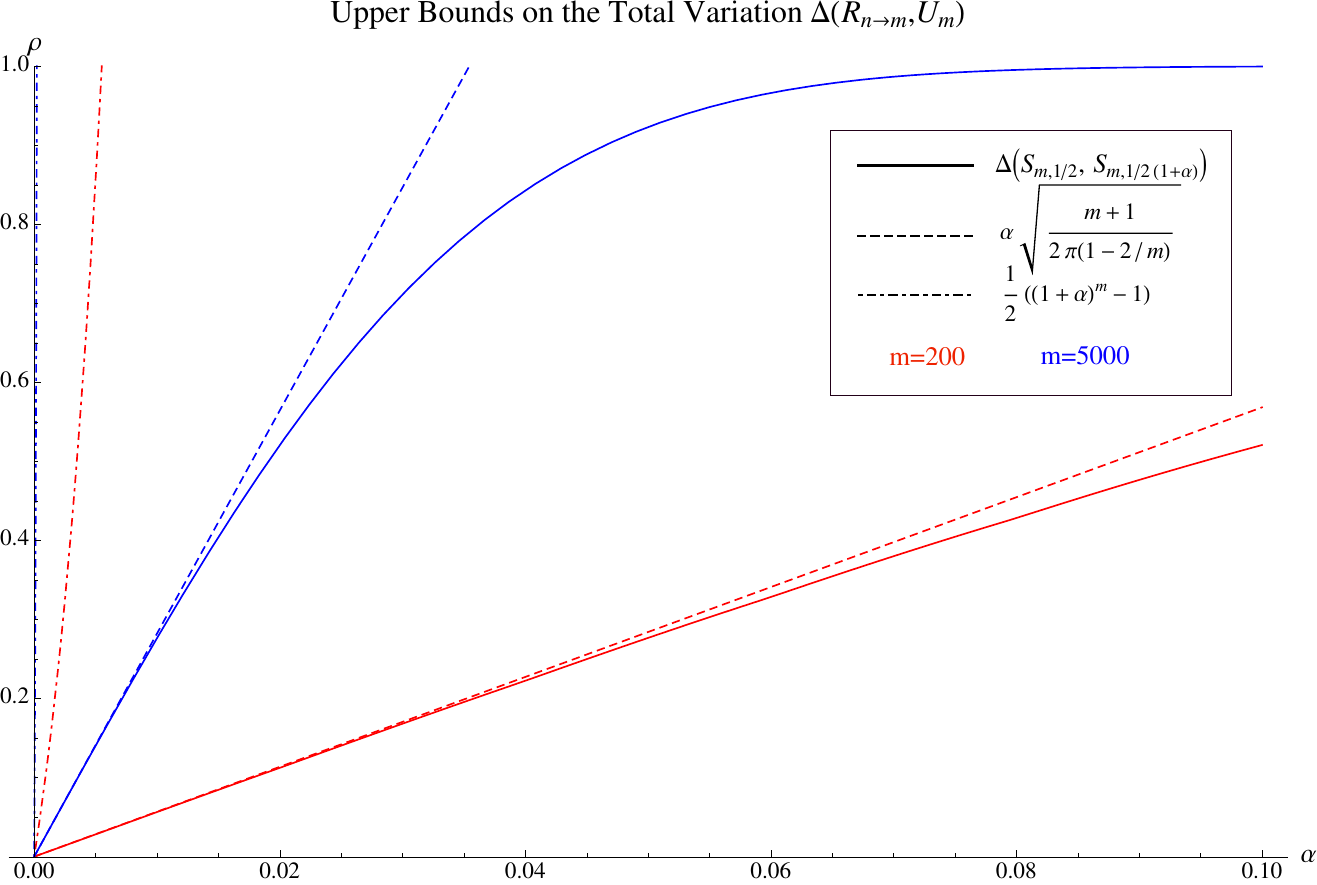}
	\caption{Plot of upper bounds on the variation between $R_{n\to m}$ and $U_m$.}\label{fig:rhoBounds}
\end{figure}

There are, unfortunately, some issues with this bound. First, as $m \to \infty$ the bound on the variation becomes infinite too. This is unreasonable as by definition we should have $\Delta(R_{n\to m},U_m) \le 1$. It only makes sense to talk about $\rho \le 1$, although in any useful situation we will require $\rho$ to be small (close to 0) so it is only of real importance that the bound is good in this situation. However,~\eqref{eqn:rhoCloseNaive} requires $\alpha$ to be significantly smaller than we really require for the two probabilities to be $\rho$ close. Even for small $\rho$ the bound is no-way near tight enough (see Figure~\ref{fig:rhoBounds}). Further, it would be instructive to examine more correctly the behaviour for large $m$ and investigate fully the nature of the relationship between $\alpha$, $m$ and $\rho$.

To rectify this and find a more reasonable bound, we carry out a finer analysis making use of the previous observation that this is the same problem as finding the variation between two binomial distributions. Let us denote a binomial probability distribution function for $n$ trials and probability of success $p$ as $S_{n,p} : \{0,\dots,n\} \to [0,1]$ where for each $A\subseteq \{0,\dots,n\}$, $$S_{n,p}(A) = \sum_{k \in A}\binom{n}{k}p^k(1-p)^{n-k}.$$ For $0\le p,p' \le 1$, we then have 
$$\Delta(S_{n,p},S_{n,p'})=\frac{1}{2}\sum_{k=0}^n \binom{n}{k}\left| p^k (1-p)^{n-k}-(p')^k (1-p')^{n-k}\right|,$$ and $$\max_{\gamma_i,\varepsilon_i}\Delta(R_{n\to m},U_m) = \Delta(S_{m,1/2(1\pm\alpha)},S_{m,1/2}).$$
\begin{Fact}\label{fact:BinomVariationSym}
	For $0\le p,p' \le 1$ we have $\Delta(S_{n,p},S_{n,p'})=\Delta(S_{n,1-p},S_{n,1-p'})$.
\end{Fact}

The total variation between two binomial distributions can be given in terms of regularised incomplete beta functions~\cite{Adell:2006vn}.
\begin{Definition}
	The \emph{incomplete beta function} is defined as $$\Beta_\ell(a,b)=\int_0^\ell u^{a-1}(1-u)^{b-1}du.$$
	For $\ell=1$ we write $\Beta_1(a,b) = \Beta(a,b)$ for the \emph{complete} beta function, or just \emph{beta function}.
	The \emph{regularised incomplete beta function} is defined as $$I_\ell(a,b)=\frac{\Beta_\ell(a,b)}{\Beta(a,b)}\raisebox{.8mm}{.}$$
\end{Definition}
\begin{Theorem}\label{thm:twoBinomVariation}
	Let $0\le p \le 1$, $q=1-p$ and $0 \le x \le q$. The total variation between two binomial distributions with probability of success $p$ and $p+x$ is
	\begin{align*}
		\Delta(S_{n,p},S_{n,p+x}) &= n\int_p^{p+x}S_{n-1,u}(\ell-1)du\\
		&= n \binom{n-1}{\ell-1}\int_p^{p+x}u^{\ell-1}(1-u)^{n-\ell}du\\
		&=I_{p+x}(\ell,n-\ell+1)-I_p(\ell,n-\ell+1),
	\end{align*}
	where $$\ceil{np}\le \ell:=\ell(n,p,x)=\ceil{\frac{-n\log\left(1-x/q\right)}{\log\left(1+x/p\right)-\log\left(1-x/q\right)}}\le \ceil{n(p+x)}.$$
\end{Theorem}
\begin{proof}
	The first line is from Adell and Jodr\'a~\cite{Adell:2006vn}. The rest follows from the well known properties of the beta functions: $\Beta_\ell(a,b)=\Beta_\ell(b,a)$ and $$\binom{n}{k}=\frac{1}{(n+1)\Beta(n-k+1,k+1)}\raisebox{.8mm}{.}$$
\end{proof}
\begin{Theorem}\label{thm:variationBoundExact}
	The total variation is bounded by
	\begin{align*}
		\Delta(R_{n\to m},U_m)&\le \Delta(S_{m,1/2},S_{m,1/2(1+\alpha)})\\
		&= I_{1/2(1+\alpha)}(\ell,m-\ell+1)-I_{1/2}(\ell,m-\ell+1),\\
		&= F(m - \ell;m,1/2-\alpha/2) - F(m - \ell;m,1/2)
	\end{align*}
	where $$\ceil{m/2}\le \ell = \ell(m,1/2,\alpha/2) = \ceil{\frac{-m\log(1-\alpha)}{\log(1+\alpha)-\log(1-\alpha)}}\le \ceil{m(1+\alpha)/2},$$
	and $$F(k;n,p) = \sum_{x=0}^kS_{n,p}(x)$$ is the cumulative distribution function for the binomial distribution.
\end{Theorem}
\begin{proof}
	This follows directly from Theorem~\ref{thm:twoBinomVariation} and Fact~\ref{fact:BinomVariationSym}. The last line follows from well known properties of the binomial distribution.
\end{proof}

\begin{figure}[h]
	\centering
	\includegraphics[scale=0.7]{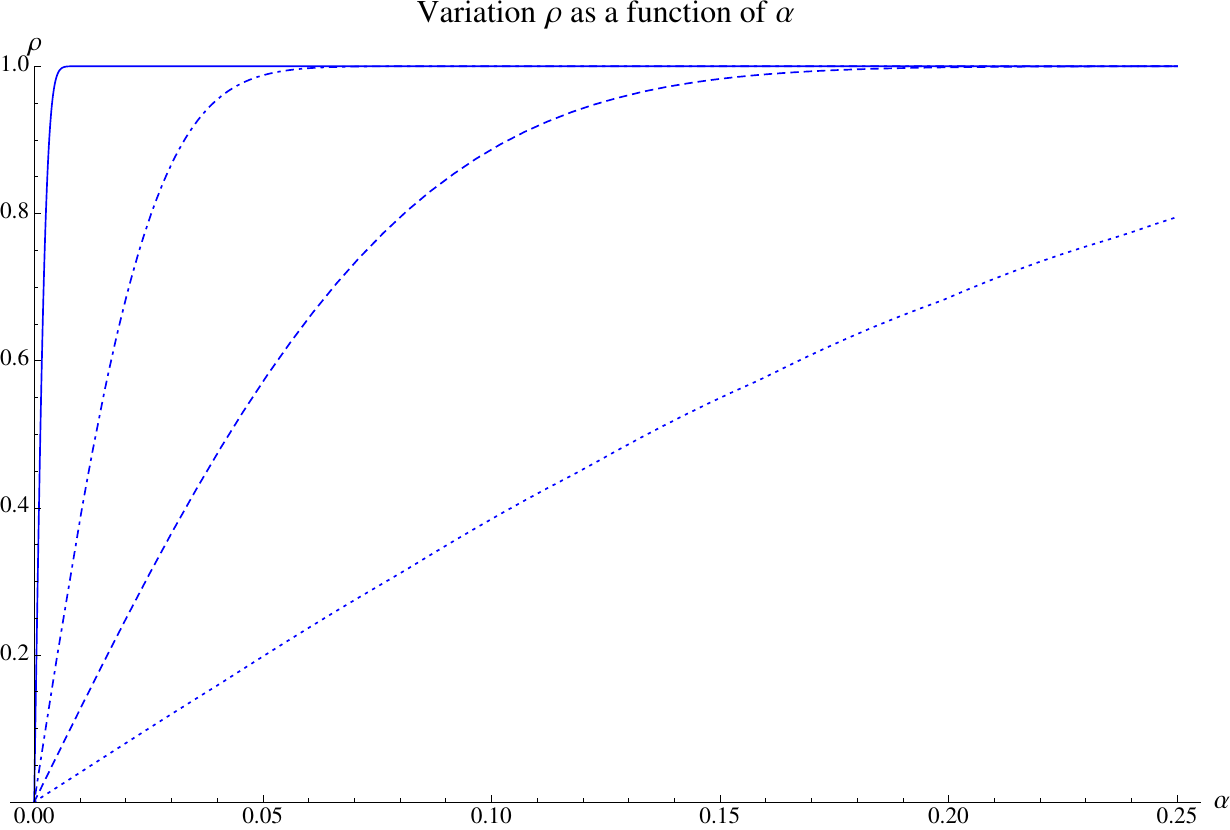}
	\caption{Plot of $\rho$ against $\alpha$ using the bound in Theorem~\ref{thm:variationBoundExact} for four values of $m$: 100 (dotted), 1,000 (dashed), 10,000 (dot-dashed) and 1,000,000 (solid).}\label{fig:rhoValphaExact}
\end{figure}

This bound is exact (under the extrema given by Lemma~\ref{lemma:maxAlpha}), and we easily verify that $\Delta(R_{n\to m},U_m) \le 1$ since $I_p(a,b) \le 1$ for all $a,b$ and $p \le 1$, and for $p' \ge p$ we have $I_{p'}(a,b) \ge I_p(a,b)$ (with equality only for $p=p'$). Unfortunately this bound on the variation has no simple closed form, so we can not easily relate $\alpha$, $m$ and $\rho$ like we did in~\eqref{eqn:rhoCloseNaive}. The shape and nature of this relationship can be seen for various values of $m$ in Figure~\ref{fig:rhoValphaExact}. In practice, with $m$ fixed and given $\rho$ it is easy to compute (with numerical methods) $\alpha$ such that $\Delta(R_{n\to m},U_m) \le \rho$. For relatively small $\rho$ however, we can find a simple and fairly good bound which is easy to work with for rough approximations. 

\begin{Theorem}\label{thm:goodRhoBound}
	Assume that $m=n/2$. Consider the probability spaces $(B^m,2^{B^m},R_{n\to m})$ and $(B^m,2^{B^m},U_m)$. For  every real $\rho$ such that $0 \le \rho < 1$, if $$\alpha \le \rho\sqrt{\frac{2\pi (1-\frac{2}{m})}{m+1}}\raisebox{.9mm}{,}$$ then $\Delta(R_{n\to m},U_m) \le \rho$.
\end{Theorem}
\begin{proof}
	We will take a first order (linear) approximation of $\Delta(S_{m,1/2},S_{m,1/2(1+\alpha)})$ around $\alpha=0$. From Theorem~\ref{thm:twoBinomVariation} and the Fundamental Theorem of Calculus we have 
	$$\Phi(\alpha) := \frac{d}{d\alpha}\Delta(S_{m,1/2},S_{m,1/2(1+\alpha)}) = m \binom{m-1}{\ell-1}2^{-m}(1+\alpha)^{\ell-1}(1-\alpha)^{m-\ell}.$$
	Since $\ell\ge \ceil{m/2}$ we have 
	$$\Phi(\alpha) \le \Phi(0),$$
	so our first order upper bound is given by 
	$$\Delta(S_{m,1/2},S_{m,1/2(1+\alpha)}) \le \alpha \Phi(0) = \alpha m \binom{m-1}{\ell-1}2^{-m}.$$
	Since the central binomial coefficient (i.e. $\binom{n}{\floor{n/2}}$) is the largest, for $k\le m-1$ we have
	$$\binom{m-1}{k} \le \binom{m-1}{\floor{\frac{m-1}{2}}} = \binom{m-1}{\ceil{\frac{m}{2}}-1},$$
	which can easily be shown by taking the two cases of $m$ odd and $m$ even. Since $\ell \ge \ceil{m/2}$ we have that 
	$$\Phi(0) \le 2^{-m}m\binom{m-1}{\ceil{\frac{m}{2}}-1} = 2^{-m}m\frac{\ceil{\frac{m}{2}}}{m}\binom{m}{\ceil{\frac{m}{2}}} = 2^{-m}\ceil{m/2}\binom{m}{\ceil{\frac{m}{2}}}\raisebox{.9mm}{.}$$
	Using the bounds given in Corollary 2.3,~\cite{Stanica:2001zr}, and writing $m=a\ceil{m/2}$ where $a\le 2$, we have
	\begin{align*}
		\binom{a\ceil{\frac{m}{2}}}{\ceil{\frac{m}{2}}} &< \frac{1}{\sqrt{2\pi \ceil{\frac{m}{2}}}}\frac{a^{m+\frac{1}{2}}}{(a-1)^{(a-1)\ceil{\frac{m}{2}}+\frac{1}{2}}}\\
		&= \frac{1}{\sqrt{2\pi \ceil{\frac{m}{2}}}}\frac{m^{m+\frac{1}{2}}}{\floor{\frac{m}{2}}^{\floor{\frac{m}{2}}+\frac{1}{2}}\ceil{\frac{m}{2}}^{\ceil{\frac{m}{2}}}}\\
		&\le \frac{1}{\sqrt{2\pi \ceil{\frac{m}{2}}}}  \frac{m^{m+\frac{1}{2}}}{\left((\frac{m}{2}+\frac{1}{2})(\frac{m}{2}-\frac{1}{2}) \right)^{\floor{\frac{m}{2}}}(\frac{m}{2}-\frac{1}{2})^{\frac{1}{2}}(\frac{m}{2})}\\
		&\le \frac{1}{\sqrt{2\pi \ceil{\frac{m}{2}}}}  \frac{2^{m+\frac{1}{2}}}{\left(1-\frac{1}{m^2} \right)^{\floor{\frac{m}{2}}} (1-\frac{1}{m})^{\frac{1}{2}}}\\
		&\le \frac{1}{\sqrt{\pi \ceil{\frac{m}{2}}}}  \frac{2^{m}}{\left(1-\frac{1}{2m} \right) (1-\frac{1}{m})^{\frac{1}{2}}}\\
		&\le \frac{2^m}{\sqrt{\pi \ceil{\frac{m}{2}}(1-\frac{2}{m})}}
		\raisebox{.9mm}{.}
	\end{align*}
	Hence, we have 
	$$\Phi(0)\le \sqrt{\frac{\ceil{\frac{m}{2}}}{\pi (1-\frac{2}{m})}} \le \sqrt{\frac{m+1}{2\pi (1-\frac{2}{m})}}\raisebox{.9mm}{.}$$
\end{proof}
This bound is much better than the  bound given in (\ref{eqn:rhoCloseNaive}), and for small $\alpha$ is extremely good. It has the desired properties that as $\alpha \to 0$, the bound on the variation tends to $0$ also. Obviously this bound is not less than one for all $\alpha$, but for small $\rho$ the bound is very good, as can be seen in Figure~\ref{fig:rhoBounds}.

\medskip

Another interesting question refers to the possibility 
of manipulating the parameter  $\alpha$ for fine calibration of the QRNG. For $R_{n\to m}$ to become closer to $U_m$ we need to make $\alpha$ smaller, but this can be done by adjusting both $\delta$ and $\beta$. As previously discussed, both are reasonable physical parameters, and which one is the most suitable (or easiest) to decrease experimentally will to a large extent depend on the QRNG set-up itself. However, adjusting $\delta$ has a larger effect on $\alpha$ than adjusting $\beta$ does, and $R_{n\to m}$ will only approach $U_m$ arbitrarily close as $\delta\to 0$, as even with $\beta=\delta$ (recall $\delta \le \beta$) we do not have $\alpha=0$ unless $\delta = 0$.

\medskip

These results can be extended  to all $m\le n/2$, although the analysis is rather ellaborated. The key difference is that in the definition of $R_{n\to m}$ in~\eqref{normalprob}  the set $VN_{n,m}^{-1}(Y)$ no longer has the same size as $Y$, so an additional summation is needed in the right hand side of~\eqref{normalprob}. However, the total variation will still be maximised under the same conditions as in Lemmata~\ref{lemma:maximiseSumAbsValues} and~\ref{lemma:maxAlpha}, and the same relation as in Theorem~\ref{thm:variationBoundExact} holds.

It is worth noting that the conditions which maximised the variation in~\eqref{eqn:uMaxConds} correspond to every $\varepsilon_i$ being the same up to a small variation $\delta$. Physically this would indicate that $p_0,p_1$ have been incorrectly stated, but that the device is actually rather accurate except for a small drift in probabilities of no more than $\delta$. Since the parameters $\varepsilon_i$ are supposed to physically account for the amount the probability is allowed to drift, which will normally be much more than the drift between individual bits (the $\gamma_i$), if the device is calibrated so that $p_0$ and $p_1$ are centred so that the $\varepsilon_i$ are distributed around them, then the variation will not be nearly as bad as in this worst case. However, the bound on the variation remains valid as it is not necessarily meaningful (or useful) to look into the physical situation under which the worst case bound is achieved.

We briefly wish to point out that other methods for dealing with independent-bit sources have been proposed. For example,  grouping bits into blocks of size $\ell$ and taking the parity of these bits for the ``normalised'' bit,  produces a string of length $n/\ell$~\cite{Vadhan:2011kx}. With this method each bit becomes unbiased exponentially fast in $\ell$.
However,  the bound in Theorem~\ref{thm:goodRhoBound} is asymptotically tighter than the corresponding bound that can be obtained by the parity method if the block size $\ell$ is fixed; if $\ell$ scales polynomially with $n$ then this method produces a better bound, but at a substantial cost to the number of bits produced~\cite[Proposition 6.5]{Vadhan:2011kx}. The reason the von Neumann normalisation outperforms the parity method is due to the fact that the bias is required to vary slowly.

\if01
\subsection*{Finite Case}

Consider a stream of \emph{independent} bits coming from a biased QRNG. Let $p_0$, $p_1$ be the probability that a bit is $0$ or $1$ respectively. Since the bits are independent, the probability of any string $x$ of length $n$ depends only on the number of $0$s, $\#_0(x)$ and $1$s, $\#_1(x)$, and the number of $1$s obeys a Binomial distribution. It is clear then that we can write $P(x) = p_0^{\#_0(x)}p_1^{\#_1(x)}$.

\begin{Theorem}
	Let $B^n=\{0,1\}^n$ and $(B^n, 2^{B^n}, P)$ with $$P(x) = p_0^{\#_0(x)}p_1^{\#_1(x)}$$ for all $x\in B^n$ be the probability space of bitstrings produced by a series of independent, biased events, i.e.\ $p_0 \neq p_1$ and $p(x_i \cap x_j) = p(x_i)p(x_j)$ for all $i,j \le n$. Let $m \le \lfloor n/2 \rfloor$ and $$B^{n\to m} = \{ x^* \mid x^* = \text{VN}(x) \text{ for some } x\in B^n \text{ and } |x^*| = m\}.$$ Then $(B^{n\to m},2^{B^{n\to m}},P^*)$, the probability space of bitstrings of length $m$ produced my von Neumann normalisation, is the uniform distribution $(B^m,2^{B^m},U)$.
\end{Theorem}
\begin{proof}
First, note that for $m \le \lfloor n/2 \rfloor$ (the maximum efficiency for normalisation), $B^{n\to m} = B^m$. This is true since for every $x^* \in B^m$ we can construct an $x \in B^n$ such that $VN(x) = x^*$ by the transformation $x^* \to f(x^*_1) \dots f(x^*_m) 0^{n-2m}$ where
\begin{equation}\label{eqn:un-norm}
	f(x^*_i) = \begin{cases} 01 & \text{if $x^*_i = 0$,}\\ 10 & \text{if $x^*_i = 1$.}\end{cases}
\end{equation}
Hence, every string of length less than or equal to $\lfloor n/2\rfloor$ is obtainable via von Neumann normalisation of a string of length $n$. 

For any $x^*\in B^m$, $P^*(x^*)$ is given by the probability of obtaining a string $x\in B^n$ such that $VN(x)=x^*$, normalised over all $x\in B^n$ such that $VN(x) \in B^m$. Let $A = \{x \mid VN(x) = x^* \}$, $N=\{x \mid VN(x) \in B^m \}$ and $D=\{00,11 \}^{\lfloor \frac{n}{2}-m\rfloor}$, then
\begin{align*}
	P^*(x^*) &= \frac{P(A)}{P(N)}\\
		&= \frac{1}{P(N)}\sum_{x\in A}P(x)\\
		&= \frac{1}{P(N)} p(01)^m \sum_{y\in D}k\cdot p(y),
\end{align*}
where $k$ is the number of (order preserving) ways to insert the blocks of $00$ and $11$ that form $y$ between the blocks of $f(x^*_i)$ to create $x$. The last line follows by dividing $x$ into blocks of two bits (a final trailing odd bit can be ignored), and using the assumption of independence to simplify the sum. Most importantly, this sum is independent of $x^*$. Since we require $P^*(B^m) = \sum_{x^* \in B^m}P^*(x^*) = 1$, and all $P^*(x^*)$ are equal, we must have $$P^*(x^*) = 1/|B^m| = 2^{-m}$$ for all $x^* \in B^m$, and hence $P^*$ is the uniform distribution $U$.
\end{proof}
Importantly, the fact that von Neumann normalisation induces the uniform distribution implies uniformity at all levels.
\fi

\section{The infinite case}

The extension of the above results to infinite sequences of bits produced by QRNGs is fairly straightforward, but forces us to address a few unexpected problems. First, we must extend the definition of the normalisation function $VN_{n,m}$ to sequences. We define $VN: B^\omega \to B^\omega \cup B^*$ as $$VN(\seq{x}=x_1\dots x_n \dots) = F(x_1 x_2)F(x_3 x_4) \cdots F(x_{2\floor{\frac{n}{2}}-1}x_{2\floor{\frac{n}{2}}})\cdots.$$ For convenience we also define $VN_n: B^\omega \to \left(\bigcup_{k\le n}B^k\right) \cup\{\lambda \}$ as $$VN_n(\seq{x}) = F(x_1 x_2)F(x_3 x_4) \cdots F(x_{2\floor{\frac{n}{2}}-1}x_{2\floor{\frac{n}{2}}}) = VN_{n,n}(x_1 \dots x_n).$$ 

Secondly, we introduce the probability space of infinite sequences as in \cite{Calude:2002fk}. Let $A_Q=\{a_1,\dots,a_Q\}$, $Q \ge 2$ be an alphabet with $Q$ elements.
We let $\mathcal{P}=\{xA_{Q}^\omega \mid x \in A_{Q}^* \} \cup \{ \emptyset \}$ and $\mathcal{C}$ be the class of all finite mutually disjoint unions of sets in $\mathcal{P}$;  the class $\mathcal{P}$ can be readily shown to generate a $\sigma$-algebra $\mathcal{M}$. 
Using Theorem~1.7 from \cite{Calude:2002fk}, the  probabilities on $\mathcal{M}$ are characterised by the functions $h : A_{Q}^* \to [0,1] $ satisfying:
\begin{enumerate}
	\item $h(\lambda) = 1$,
	\item $h(x) = h(x_{a_{1}}) + \dots +  h(x_{a_{Q}})$, for all $x \in A_{Q}^*$.
\end{enumerate}

If $Q=2$ so $A_{2}=B$, and for $x\in B^n$ we take $h(x) = P_n(\{x\})$ with $P_n$ as defined in Fact~\ref{prop:finiteProbSpace}, then the above conditions are satisfied. This induces our probability measure $\mu_P$ on $\mathcal{M}$, which satisfies $\mu_P(XB^\omega) = P_n(X)$ for $X\subseteq B^n$. 
Hence the suitable extension of the finite case probability space to infinite generated sequences is the space $(B^\omega, \mathcal{M},\mu_P)$.  In the special case when $p_{0}=p_{1}$ we get the
Lebesgue probability $\mu_{P_{L}}(XB^\omega) = \sum_{x\in X}2^{-|x|}$.

 In general, if $Q\ge 2$, $p_i \ge 0$ for $i=1,\dots,Q$ are  reals in [0,1] such that $\sum_{i=1}^Q p_i =1$, we can take $h_{Q}(x) = p_1^{\#_{a_{1}}(x)}\dots p_Q^{\#_{a_{Q}}(x)}$ ($\#_{a_{i}}(x)$ is the number of occurrences of $a_{i}$ in $x$) to obtain the probability space  $(A_{Q}^{\omega}, \mathcal{M},\mu_{P_{Q}})$
 in which $\mu_{P_{Q}}(xA_{Q}^\omega ) = h_Q(x)$, for all $x \in A_{Q}^*$.

The first result notes that there exist sequences $\seq{x} \in B^\omega$ such that $VN(\seq{x})\in B^*$. In fact every string can be produced via von Neumann normalisation from a suitable sequence.

\begin{Theorem}\label{col}
	For every  string $y \in B^*$ there exists an uncountable set $R \subset B^\omega$ of $\mu_P$ measure zero such that for all $\seq{x} \in R$, $VN(\seq{x})=y$.
\end{Theorem}
\begin{proof}
	Let $y = y_1 \dots y_n \in B^*$ and $D=\{00,11\}$, the two-bit blocks which are deleted by von Neumann normalisation and $y'=f(y_1) \dots f(y_n)$. Then every sequence $\seq{x} \in y' D^\omega $ satisfies $VN(\seq{x})=VN_{2n}(\seq{x})VN(x_{2n+1}x_{2n+2}\dots)=y$ since $VN_{2n}(\seq{x})=VN_{2n,2n}(y')=y$ and for all $\seq{z}\in D^\omega$ we have $VN(\seq{z})=\lambda$. 
	Obviously, the set  $R= y' D^\omega$ is uncountable and has $\mu_P$ measure zero as the set of Borel normal sequences has measure one~\cite{Calude:2002fk}.
\end{proof}
\begin{Corollary}\label{VNfiniteProbzero}
	The set $Q=\{\seq{x} \in B^\omega \mid VN(x) \in B^* \}$ has $\mu_P$ measure zero.
\end{Corollary}
\begin{proof}
	We simply note that the union of countably many measure zero sets also has measure zero.
\end{proof}

It is interesting to note that the ``collapse'' in the generated sequence produced by von Neumann normalisation in Theorem~\ref{col}  is not due to computability properties of the sequence.  In particular, there are random sequences that collapse to any string, so  to strings which are not Borel normal. 

\medskip

In the following we need a measure-theoretic characterisation of random sequences, so we present a few facts from constructive  topology and probability.

Consider  the
   compact topological space $(A_{Q}^{\omega}, \tau)$ in which the basic open sets are  the sets
$wA_{Q}^{\omega}$, with $w \in A_{Q}^{*}$. Accordingly, an open set $G \subset 
A_{Q}^{\omega}$ is of the form $G=VA_{Q}^{\omega}$, where $V \subset A_{Q}^{*}$.

From now on we assume that the reals $p_{i}, 1 \le i \le Q$ which define the probability $\mu_{P_{Q}}$ are all computable.  A constructively open set
$G \subset A_{Q}^{\omega}$ is an open set $G=VA_{Q}^{\omega}$ for which
$V \subset A_{Q}^{*}$ is computably enumerable (c.e.).
A constructive sequence of constructively open
 sets,  c.s.c.o. sets for short,  is a
sequence $(G_{m})_{m \geq 1}$ of constructively open sets $G_{m}=V_{m}A_{Q}^{\omega}$ such
that there exists a c.e.\  set $X \subset A_{Q}^{*} \times \N$ with
$V_{m}=\{x \in A_{Q}^{*} \mid (x,m) \in X\},$
for all natural $m \geq 1$.
A constructively null set $S \subset A_{Q}^{\omega}$ is a set for which there  exists a
c.s.c.o. sets $(G_{m})_{m \geq 1}$ with 
$S \subset \bigcap_{m \geq 1} G_{m},$
$ \mu_{P_Q} (G_{m}) \le 2^{-m}$. A sequence $\seq{x}\in A^{\omega}_{Q}$ is  random in the probability space $(A_{Q}^{\omega}, \mathcal{M},\mu_{P_{Q}})$
if  $\seq{x}$ is not contained in any constructively null set in $(A_{Q}^{\omega}, \mathcal{M},\mu_{P_{Q}})$. For the case of the
Lebesgue probability $\mu_{P_{L}}$ 
 the measure-theoretic characterisation of random sequences holds true:  $\seq{x}$ is  random if and only if $\seq{x}$ is not contained in any constructively null set of  $(A_{Q}^{\omega}, \mathcal{M},\mu_{P_{L}})$
~\cite{Martin-Lof:1966kx,Calude:2002fk}.

\medskip

We continue with another instance in which von Neumann normalisation decreases randomness.

\begin{Proposition}
\label{1/2}
There exist (continuously many) infinite 1/2-random sequences $\seq{x} \in B^\omega$ such that $VN(\seq{x}) = 000\dots 00\dots$.
\end{Proposition}
\begin{proof} Consider a random sequence $\seq{x} = x_{1}x_{2} \dots x_{n}\dots$ and construct the sequence $\seq{x'}=0x_{1}0x_{2} \dots 0x_{n}\dots$. Clearly, $\seq{x'}$ is 1/2-random, but $VN(\seq{x'})=  000\dots 00\dots$ because there exist infinitely many 1's in  $\seq{x}$.
\end{proof}

\medskip

We follow this with instances for which the converse is true: von Neumann normalisation conserves or increases randomness.

\begin{Proposition}
\label{1}
There exist (continuously many) infinite 1/2-random sequences $\seq{x} \in B^\omega$ such that $VN(\seq{x})$ is random.
\end{Proposition}
\begin{proof} Consider a random sequence $\seq{x} = x_{1}x_{2} \dots x_{n}\dots$ and construct the sequence $\seq{x'}=x_{1}\bar{x}_{1}x_{2}\bar{x}_{2} \dots x_{n}\bar{x}_{n}\dots$. Clearly, $\seq{x'}$ is 1/2-random and $VN(\seq{x'})=  \seq{x}$.
\end{proof}

\noindent {\bf Comment.} Both Proposition~\ref{1/2} and~\ref{1} are true for the more general case of $\varepsilon$-random sequences, where $0< \varepsilon < 1$ is computable.

\medskip

We briefly note that in the definition of Borel normality it does not matter if we count the number of non-overlapping occurrences of each string of length $m$, $N_i^m(y)$ as defined in Section~\ref{sec:notation}, or the number of overlapping occurrences, $\mathcal{N}_i^m(y)$~\cite{Kuipers:1974fk}.
One of the main results of this section, presented in Theorem~\ref{bnormal}, is the following: Borel normality is invariant under von Neumann normalisation.

\if01
\begin{Theorem}\label{thm:VN_inf-finite_notBN}
	Every $\seq{x}=x_1 x_2 \dots \in B^\omega$ such that $VN(\seq{x}) \in B^*$ is not Borel normal.
\end{Theorem}
\begin{proof}
	The number of occurrences (as defined in \cite{Calude:2002fk}) of the strings $01$ or $10$ in $\seq{x}$, $\mathcal{N}_2^2(\seq{x}) + \mathcal{N}_3^{2}(\seq{x}) = |VN(\seq{x})|$, and is thus finite. Hence 
	$$\lim_{n\to \infty}\frac{\mathcal{N}_2^2(\seq{x}(n))}{\lfloor \frac{n}{2}\rfloor}=0\neq 2^{-2}.$$
	Thus we conclude $\seq{x}$ is not Borel $2$-normal and hence not Borel normal. Indeed this can easily be extended to show $\seq{x}$ is not Borel $m$-normal for any $m\ge 2$.
\end{proof}
\begin{Corollary}
	Every $\seq{x} \in B^\omega$ such that $VN(\seq{x}) \in B^*$ is not  random.
\end{Corollary}
\begin{proof}
	This follows immediately, since all algorithmically random sequences are Borel normal.
\end{proof}
\fi

\begin{Theorem}\label{bnormal}
	Let $\seq{x} \in B^\omega$ be Borel normal in $(B^\omega,\mathcal{M},\mu_{P_L})$. Then $VN(\seq{x})$ is also Borel normal in $(B^\omega,\mathcal{M},\mu_{P_L})$.
\end{Theorem}
\begin{proof}
	Note that $VN(\seq{x}) \in B^\omega$ because $\seq{x}$ contains infinitely many occurrences 
	of 01 on even/odd positions. Let $D=\{00,11\}$, $\seq{x}^*(n) = VN_{n,n}(\seq{x}(n))$, $n'=|\seq{x}^*(n)|$. We have
	\begin{align*}
		\lim_{n'\to \infty}\frac{N_i^m(\seq{x}^*(n))}{n'} &= \lim_{n'\to \infty}\left(\frac{n}{n'}\right) \left(\frac{N_i^m(\seq{x}^*(n))}{n}\right)\raisebox{.9mm}{,}
	\end{align*}
	but as $n\to \infty$, $n' \to \infty$. We thus have
	\begin{align*}
		\lim_{n'\to \infty}\frac{n'}{n} &= \lim_{n'\to \infty}\frac{N_{0}^1(\seq{x}^*(n)) + N_{1}^1(\seq{x}^*(n))}{n}\\
		&= \lim_{n\to \infty}\frac{\mathcal{N}_{01}^2(\seq{x}(n)) + \mathcal{N}_{10}^2(\seq{x}(n))}{\floor{n/2}}\\
		&= 2^{-1}
	\end{align*}
	by the normality of $\seq{x}$. The number of occurrences of each $i=i_1\dots i_m\in B^m$ in $\seq{x}^*(n)$ is the number of occurrences of $i'=f(i_1)y_1f(i_2)\dots y_{m-1}f(i_m)$ in $\seq{x}(n)$, summed over all $y_1, \dots, y_{m-1} \in D^*$. Viewing $i'$ as a string over $\{00,01,10,11\}$ we have:
	\begin{align*}
		\lim_{n'\to \infty}\frac{N_i^m(\seq{x}^*(n))}{n} &= \lim_{n\to \infty}\frac{\sum_{y_1,\dots,y_{m-1}}N_{i'}^{|i'|}(\seq{x}(n))}{n}\\
		&=\sum_{y_1\in D^*}\sum_{y_2\in D^*}\cdots \sum_{y_{m-1}\in D^*} 2^{-2|i'|}\\
		&=\sum_{|y_1|=0}^\infty 2^{|y_1|}\sum_{|y_2|=0}^\infty 2^{|y_2|}\cdots \sum_{|y_{m-1}|=0}^\infty 2^{|y_{m-1}|}2^{-2|i'|}\\
		&=2^{-2m}\sum_{|y_1|=0}^\infty 2^{-|y_1|}\sum_{|y_2|=0}^\infty 2^{-|y_2|}\cdots \sum_{|y_{m-1}|=0}^\infty 2^{-|y_{m-1}|}\\
		&= 2^{-2m}2^{m-1}\\
		&= 2^{-(m+1)}.
	\end{align*}
	Hence, both limits exist and we have 
	\begin{align*}
			\lim_{n'\to \infty}\frac{N_i^m(\seq{x}^*(n))}{n'} &= \lim_{n'\to \infty}\left(\frac{n}{n'}\right) \left(\frac{N_i^m(\seq{x}^*(n))}{n}\right)\\
			&= \frac{\lim_{n'\to \infty}\frac{N_i^m(\seq{x}^*(n))}{n}}{\lim_{n'\to \infty}\frac{n'}{n}}\\
			&= \frac{2^{-(m+1)}}{2^{-1}}\\
			&= 2^{-m}.
	\end{align*}
	Since this holds for all $m,i$ we have that $VN(\seq{x})$ is Borel normal.
\end{proof}


	Let $A_Q=\{a_1,\dots,a_Q\}$, $Q \ge 3$. Let $\sum_{i=1}^Q p_i =1$ where $p_i \ge 0$ for $i=1,\dots,Q$ and  $(A_{Q}^{\omega}, \mathcal{M},\mu_{P_{Q}})$ be the probability space  defined by the probabilities $p_i$. Let $A_{Q-1}=\{a_1,\dots,a_{Q-1}\}$ and $(A_{Q-1}^{\omega}, \mathcal{M},\mu_{P^{T}_{Q-1}})$ be the probability space  defined by the probabilities 
	$$p_i^{T} = p_i\left(1+\frac{p_Q}{\sum_{j=1}^{Q-1}p_j} \right) = \frac{p_{i}}{1-p_{Q}}\raisebox{.9mm}{,}$$
	with $1\le i \le Q-1$. Let $T:A_Q^* \to A_{Q-1}^*$  be the monoid morphism defined by $T(a_i) = a_i$ for $1\le i \le Q-1$, $T(a_Q) = \lambda$;  $T(x)=T(x_1)T(x_2)\cdots T(x_n)$ for $x\in A_Q^n$.  As $T$ is prefix-increasing we naturally extend 
	$T$ to sequences to obtain the function $T:  A_Q^\omega \to A_{Q-1}^\omega$ given by  $T(\seq{x})=\lim_{n\to\infty}T(\seq{x}(n))$ for $\seq{x} \in A_Q^\omega$. 
	
\begin{Lemma}\label{TMeasInv}
The transformation $T$ is $(\mu_{P_Q},\mu_{P^{T}_{Q-1}})$--preserving, i.e.\ for all $w\in A_{Q-1}^*$ we have $ \mu_{P_Q} \left(T^{-1}(w A_{Q-1}^\omega)\right) = \mu_{P^{T}_{Q-1}}\left(wA_{Q-1}^\omega\right)$.
\end{Lemma}
\begin{proof} Take $w=w_{1}\dots w_{m}\in A_{Q-1}^\omega$. We have:
\begin{align*}
		\mu_{P_Q} \left(T^{-1}(w A_{Q-1}^\omega)\right)  &= \mu_{P_Q}\left( \{\seq{x} \in A^{\omega}_{Q} \mid w  \sqsubset T(\seq{x})\}\right)\\	
		&= \mu_{P_Q}\left\{a_{Q}^{i_{1}}w_{1}a_{Q}^{i_{2}}w_{2} \dots a_{Q}^{i_{m}}w_{m}\seq{z}\mid \seq{z}\in 
		A^{\omega}_{Q}\right\}\\
		&= \sum_{i_{1}, \dots ,i_{m} =0}^{\infty} h_{Q}\left(a_{Q}^{i_{1}}w_{1}a_{Q}^{i_{2}}w_{2} \dots a_{Q}^{i_{m}}w_{m}\right) \\
		&= \sum_{i_{1}, \dots ,i_{m} =0}^{\infty} h_{Q-1}(w) \cdot p_{Q}^{i_{1} + \dots +i_{m} }\\
		&= h_{Q-1}(w) \cdot \frac{1}{1-p_{Q}}\\
		& = h^{T}_{Q-1} (w)\\
		& = \mu_{P^{T}_{Q-1}}\left(wA_{Q-1}^\omega\right).
	\end{align*}

\end{proof}

\begin{Proposition}\label{changebase} If $\seq{x}\in A_{Q}^\omega$ is   random in  $(A_{Q}^{\omega}, \mathcal{M},\mu_{P_{Q}})$ and $T$ is the transformation defined in Lemma~\ref{TMeasInv}, then
$T(\seq{x})$ is random in $(A_{Q-1}^{\omega}, \mathcal{M},\mu_{P^{T}_{Q-1}})$.
\end{Proposition}
\begin{proof}
	We generalise a result in~\cite{Calude:2001uq} stating that, for the Lebesgue probability, measure-preserving transformations preserve randomness. Assume that $\seq{x}$ is   random in  $(A_{Q}^{\omega}, \mathcal{M},\mu_{P_{Q}})$ but $T(\seq{x})$ is not random in in $(A_{Q-1}^{\omega}, \mathcal{M},\mu_{P^{T}_{Q-1}})$, i.e.\  there is a constructive null set $R= (G_{m})_{m\ge 1}$ containing $T(\seq{x})$. Assume that $G_{m}= X_{m}A^{\omega}_{Q-1}$, where $X_{m}
	\subset A^{\omega}_{Q-1}$ is c.e.\ and  has the measure $\mu_{P^{T}_{Q-1}}(X_{m}A^{\omega}_{Q-1})$ smaller than $2^{-m}$.
	Define $S_{m}= T^{-1}(X_{m}A^{\omega}_{Q-1}) \subset A^{\omega}_{Q}$ and note that $S_{m} $ is open because it is equal to  $\bigcup_{w\in X_{m}} V_{w}A^{\omega}_{Q}$ with $V_{w}= \{ v \in A^{\omega}_{Q} \mid w \sqsubset T(v) \}$ and, using Lemma~\ref{TMeasInv}, has the measure smaller than $2^{-m}$:
	\begin{align*}
	\mu_{P_{Q}}(S_{m}) & = \mu_{P_{Q}}\left(\bigcup_{w\in X_{m}} V_{w}A^{\omega}_{Q}\right)\\
	& \le \sum_{w\in X_{m}} \mu_{P_{Q}}\left(V_{w}A^{\omega}_{Q}\right)\\
	& =  \sum_{w\in X_{m}} \mu_{P_{Q}} \left(T^{-1}\left(wA^{\omega}_{Q-1}\right)\right)\\
	& = \mu_{P^{T}_{Q-1}}\left(X_{m}A^{\omega}_{Q-1}\right)\\
	& \le 2^{-m}.
	\end{align*}
We have proved that $\seq{x}$ is not random in  $(A_{Q}^{\omega}, \mathcal{M},\mu_{P_{Q}})$, a contradiction.
\end{proof}

Let us define $VN^{-1} : 2^{B^*} \to 2^{B^*}$ for $x=x_1 \dots x_m \in B^m$ as 
\begin{align*}
	VN^{-1}(x)&=\{y \mid y=u_1 f(x_1)u_2 \dots u_m f(x_m)u_{m+1}v \text{ and } \\
	& \phantom{==} u_i \in \{00,11\}^* \text{ for $1 \le i \le m$}, v \in B\cup\{\lambda\}\}\\
	&= \bigcup_{n=0}^\infty VN_{n+2m,m}^{-1}(x),
\end{align*}
and for $X\subseteq B^*$ as $$VN^{-1}(X)=\bigcup_{x\in X}VN^{-1}(x).$$
For all $x \in B^*$ and $\seq{y}\in VN^{-1}(x)B^\omega$ we then have $x \sqsubset VN(\seq{y})$.

\if01
Let us define $$X(x^*)=\{x\mid x=d_1 f(x_1^*)d_2 \dots d_n f(x_n^*) \text{ and } d_i \in D^* \text{ for $1 \le i \le n$}\}$$ for $x^* \in B^*$. 
The natural extension of $VN_{n,m}^{-1}$ is the preimage of $VN(xB^\omega)$, $VN^{-1} : 2^{B^\omega} \to 2^{B^\omega}$, defined for $x^*\in B^*$ as \begin{align*}
	VN^{-1}(x^*B^\omega)&=\{\seq{x} \in B^\omega \mid VN(\seq{x}) \in x^* B^\omega \}\\
	&= \{X(x^*)B^\omega \}.
\end{align*}
\fi

For the cases that $VN(\seq{x}) \in B^\omega$, the probability space $(B^\omega, \mathcal{M}, \mu_{P_{VN}})$ induced by von Neumann normalisation is endowed with the measure $\mu_{P_{VN}}$. The measure $\mu_{P_{VN}}$ is defined on the sets $xB^\omega$ with $x\in B^*$ by
$$\mu_{P_{VN}}(xB^\omega) = \frac{\mu_P(VN^{-1}(x)B^\omega)}{\mu_P(VN^{-1}(B^{|x|})B^\omega)}\raisebox{.9mm}{.}$$
By noting that $VN^{-1}(B^{|x|}) \subset VN^{-1}(B^*)$ it is clear to see that $\mu_{P_{VN}}$ satisfies the Kolmogorov axioms for a probability measure. While the set $VN^{-1}(B^{|x|})$ contains sequences for which normalisation produces a finite string, from Corollary~\ref{VNfiniteProbzero} we know that the set of such sequences have measure zero, so the definition of $\mu_{P_{VN}}$ is a good model of the target probability space.
We thus arrive at the key result that (measure-theoretical) randomness is invariant under von Neumann normalisation.

\begin{Theorem} \label{infrand}
	Let $\seq{x} \in B^\omega$ be   random in  $(B^\omega, \mathcal{M},\mu_P)$. Then $VN(\seq{x})\in B^\omega$   is also   random in  $(B^\omega, \mathcal{M},\mu_{P_{VN}})$.
\end{Theorem}
\begin{proof} We write the   random  sequence $\seq{x}$ as $\seq{x} = x_{1}x_{2}\dots x_{n}\dots = (x_{1}x_{2})
\dots (x_{2n-1}x_{2n}) \dots \in \{00,01,10,11\}^{\omega}$. Renaming $a=00, A=01, B=10, b=11$ and consistently
deleting first all occurrences of $a$ we get a   random  sequence $\seq{x_{\it A,B,b}}$  on the alphabet $\{A,B,b\}$, then 
deleting  all occurrences of $b$ we get a   random  sequence $\seq{x_{\it A,B}}$  on the alphabet $\{A,B\}$. The result follows from the fact that
$VN(\seq{x}) = \seq{x}_{0,1}$ and Proposition~\ref{changebase} stating that $\seq{x_{\it A,B}}$ is   random.
\end{proof}

\begin{Corollary} If $\seq{x} \in B^\omega$ is   random in  $(B^\omega, \mathcal{M},\mu_P)$ then $VN(\seq{x})$ is  Borel normal in $(B^\omega, \mathcal{M}, \mu_{P_{VN}})$.
\end{Corollary}
\begin{proof}
From Theorem~\ref{infrand} it follows that  $VN(\seq{x})$ is Borel normal  provided $\seq{x}$ is   random  \cite{Calude:2002fk}. 
\end{proof}

\begin{Theorem}
	The probability space $(B^\omega, \mathcal{M}, \mu_{P_{VN}})$ induced by von Neumann normalisation is the uniform distribution $(B^\omega, \mathcal{M}, \mu_{P_L})$, where $\mu_{P_L}$ is the Lebesgue measure.
\end{Theorem}
\begin{proof}
	By Lemma~\ref{TMeasInv} von Neumann normalisation is measure preserving, so for $x\in B^*$ we have
	\begin{align*}
		\mu_{P_{VN}}(x B^\omega) &= \mu_P(VN^{-1}(x)B^\omega)\\
			&=p_0^{|x|} p_1^{|x|}\sum_{d_i \in D^*}p_0^{\#_0(d_1 \dots d_{|x|})} p_1^{\#_1(d_1 \dots d_{|x|})}.
	\end{align*}
	The key point, as in the finite case, is that this only depends on $|x|$ not $x$ itself. By using the fact that for any $n$, $\sum_{x\in B^n}\mu_{P_{VN}}(x B^\omega) = 1$, we have $$\mu_{P_{VN}}(xB^\omega) = 2^{-|x|}$$ for all $x \in B^*$, and hence $\mu_{P_{VN}} = \mu_{P_L}$, the Lebesgue measure.
\end{proof}
\if01
\begin{proof}
	Let $D=\{00,11\}$ and note that for all $d\in D^*$, $VN_{|d|,|d|}(d)=\lambda$. Consider a string $x^* \in B^*$: the set $$X(x^*)=\{x=d_1 f(x_1^*)d_2 \dots d_n f(x_n^*)\mid d_i \in D^* \}$$ is such that for every sequence $\seq{x}\in B^\omega$, the first $n$ bits of $VN(\seq{x})$ are exactly $x^*=x_1^*,\dots,x_n^*$ if and only if $\seq{x}\in X(x^*)B^\omega$. Then
	\begin{align*}
		\mu_{P_{VN}}(x B^\omega) &= \frac{\mu_P(X(x)B^\omega)}{\mu_P(B^\omega)}\\
			&=\sum_{x\in X}P_n(\{x\})\\
			&=p_0^n p_1^n\sum_{d_i \in D^*}p_0^{\#_0(d_1 \dots d_n)} p_1^{\#_1(d_1 \dots d_n)}.
	\end{align*}
	The key point, as in the finite case, is that this only depends on $|x|$ not $x$ itself. By using the fact that for any $n$, $\sum_{x\in B^n}\mu_{P_{VN}}(x B^\omega) = 1$, we have $$\mu_{P_{VN}}(xB^\omega) = 2^{-|x|}$$ for all $x \in B^*$, and hence $\mu_{P_{VN}} = \mu_{P_L}$, the Lebesgue measure.
\end{proof}
\fi
This can easily be extended from the case when $VN(\seq{x})$ is infinite, to the case in which it is finite. To do so, note that if $\seq{y}\in B^\omega$ and $VN(\seq{x})=y \in B^n$, then the probability space induced by von Neumann normalisation is $(B^n,2^{B^n},P_n^*)$.
We then have $$P_n^*(x) = \frac{\mu_P(VN^{-1}(x)D^\omega)}{\mu_P(VN^{-1}(B^n) D^\omega)}\raisebox{.9mm}{,}$$ and since the denominator is constant for all $x \in B^n$, we can proceed as for above, and $P_n^* = U_n$ as desired.

\begin{Theorem}
	The set $\{\seq{x} \in B^\omega \mid VN(\seq{x}) \in B^{*} \text{ or  } VN(\seq{x}) \in B^{\omega} \text{ is computable }\}$ has measure zero with respect to the probability space $(B^\omega,\mathcal{M},\mu_P)$.
\end{Theorem}

\begin{proof}
By  Theorem~\ref{infrand} we deduce that $$\{\seq{x} \in B^\omega \mid VN(\seq{x}) \in B^{\omega} \text{ is computable }\} \subset \{ \seq{x} \in B^\omega \mid \seq{x} \text{ is not random in $(B^\omega,\mathcal{M},\mu_P)$} \},$$ which has measure zero~\cite{Martin-Lof:1966kx}. To complete the proof, note that we know from Corollary~\ref{VNfiniteProbzero} that the set $\{\seq{x} \in B^\omega \mid VN(\seq{x}) \in B^*\}$ also has measure zero.

\end{proof}

\section{Role of probability spaces for QRNGs}

The treatment of QRNGs as entirely probabilistic devices is grounded purely on the probabilistic treatment of measurement in quantum mechanics which originated with Born's decision to ``give up determinism in the world of atoms''~\cite{Born:1926aa}, a viewpoint which has become a core part of our understanding of quantum mechanics. This is formalised by the Born rule, but the probabilistic nature of \emph{individual} measurement is nonetheless postulated and tells us nothing about \emph{how} the probability arises. Along with the assumption of independence this allows us to predict the probability of \emph{successive} events, as we have done.

No-go theorems such as the Kochen-Specker Theorem~\cite{Kochen:2017aa} tell us something stronger: if we assume non-contextuality (i.e.\ that the result of an observation is independent of the compatible observables are co-measured alongside it~\cite{Bell:1966uq,Heywood:1983kx}) then there can, in general, be no pre-existing definite values prescribable to certain sets of measurement outcomes in dimension three or greater Hilbert space. In other words, the randomness is not due to ignorance of the system being measured; indeed, since there are in general no definite values associated with the measured observable it is surprising there is an outcome at all~\cite{Svozil:2004aa}. While this does not answer the question as to where the randomness arises from, it does tell us something stronger than the Born Rule does. In~\cite{Calude:2008aa} it is shown that every infinite sequence produced by a QRNG is (strongly) incomputable. In particular, this implies that it is \emph{impossible} for a QRNG to output a computable sequence. The set of computable numbers has measure zero with respect the probability space of the QRNG, but the impossibility of producing such sequence is much stronger than, although not in contradiction with, the probabilistic results.

In the finite case every string is, of course, obtainable, and we would expect the distribution to be that predicted by the probability space derived from the Born Rule. However, the infinite case has something to say here too. We can view any finite string produced by a QRNG as the initial segment of an infinite sequence the QRNG would produce if left to run indefinitely. For any infinite sequence produced by the QRNG, it is impossible to compute the value of \emph{any bit} before it is measured~\cite{Abbott:aa}; in the finite case this means there is no way to provably compute the value of the next bit before it is measured. In light of value indefiniteness this is not unexpected, but nonetheless gives mathematical grounding to the postulated unpredictability of each individual measurement, as well as the independence of successive measurements---indeed we can rule out any computable causal link within the system which may give rise to the measurement outcome.

The results we have presented in this paper, however, describe thoroughly the distribution of strings/sequences produced by QRNGs. With the distributions known we can create more intelligent tests of the quality of output of a QRNG~\cite{Calude:2010aa}. Current statistical tests for analysing RNGs are designed with pseudo-RNGs in mind, and are not necessarily the best way to test the quality of QRNGs. The effects of normalisation on strings generated by QRNGs can help us design QRNGs which are more robust to experimental imperfection and exhibit the desired behaviour. It will further aid in developing new normalisation techniques designed to produce the expected (ideal) theoretical distribution even in the absence of experimental imperfections.

\if01
\begin{itemize}
	\item Working with the probability spaces as we have done here to describe the output of the QRNG assumes a probabilistic/statistical outcome for each measurement given by the Born rule. It is independent of \emph{how} the probability arises, in particular it makes no reliance or use of value indefiniteness or complimentarily.
	\item By the computability results, we know that in the infinite case this is not sufficient to describe the possible outcomes of the QRNG. This is precisely because of value indefiniteness and these properties which are beyond the probability space treatment.
	\item The sets of forbidden outcomes  have measure zero, so this does not contradict the probabilistic treatment. But it says something stronger: there are measure zero events not only have probability zero but \emph{can not} be the outcome of a QRNG.
	\item In the finite case this is therefore not an issue, the probabilistic treatment is valid. However, the computability results on infinite sequences tell us about the asymptotic properties of these sequences, and the ``no bit is computable'' theorem tells us something strong about the source and the inability to (via deterministic means) prove what the next bit will be. The best we can do is the guess based on the probability spaces.
	\item Results in this paper primarily based around the form of the distribution we expect from a QRNG. This allows us to analyse expected properties of strings from QRNG in comparison the classical source. For any test, we can create a classical RNG which will give the same expected measure on the test as for the QRNG (I think?). This allows us to design robust QRNG more easily, and QRNG designs can be evaluated based on the theory, and good tests for the quality of their output developed.
\end{itemize}
\fi

\section{Conclusions}

 The  analysis developed in this paper
 involves the probability spaces of the source  and output of a QRNG
 and 
  the effect von Neumann normalisation has on these spaces.

  In the ``ideal case'',  the  von Neumann normalised output  of an independent constantly biased QRNG is the probability space of the uniform distribution (un-biasing). This result is true for both for  finite strings and for the infinite sequences produced  by  QRNGs (the QRNG runs indefinitely in the second case). 

For a real-world QRNG in which the bias, rather than holding steady, drifts slowly, we  evaluated  the speed of drift required to be maintained by the source distribution to guarantee that the output distribution is arbitrarily close  to the uniform distribution. It is an {\em open question} to study the quality of von Neumann normalisation in the  more realistic
 case when, instead of the bits being independent, the probability for each bit depends on a finite number of preceding bits (for example, because of the high bit-rate of the experiment). 
Note that Blum's algorithm~\cite{Blum:1986fv} assumes a Markov-type correlation, which cannot be assumed for a QRNG certified by
value indefiniteness~\cite{Abbott:2010fk}.
 
We have also examined the effect von Neumann normalisation has on various properties of infinite sequences. In particular, 
Borel normality and (algorithmic) randomness are invariant under normalisation, but for $\varepsilon$-random sequences with $0< \varepsilon < 1$, normalisation can both decrease or increase the randomness of the source. It is an {\it open question} whether von Neumann normalisation preserves randomness and  Borel normality for finite strings.

Finally, we reiterate that a successful application of von Neumann normalisation---in, fact, any un-biasing transformation---does exactly what it promises, {\it un-biasing},  one  (among infinitely many) symptoms of randomness; it will not produce ``true'' randomness.

\section*{Acknowledgment}
We thank Karl Svozil and Marius Zimand for many discussions and suggestions on the topics of the paper, as well as the anonymous referees for suggestions which improved the presentation of the paper.

\if01
\section*{Some notes for us}

\begin{itemize}
	\item Problem: Assume $H(x_{1}\dots x_{n}) \ge n-c$ then $VN_{n,m}(x_{1}\dots x_{n}) = y_{1} \dots y_{m}$. 
	Does there exist $c_{n,m}$ such that $H(y_{1} \dots y_{m}) \ge m - (c+c_{n,m})$?
\end{itemize}
\fi
\bibliographystyle{abbrv}
\bibliography{alastairBib}

\end{document}